\documentclass{aa}  
\usepackage{graphicx}
\usepackage{mathrsfs}
\usepackage[varg]{txfonts}
\usepackage{booktabs}
\usepackage[breaklinks, colorlinks, citecolor=blue]{hyperref}
\usepackage{mhchem}
\usepackage{tcolorbox}

\newcommand{\be}{\begin{equation}}
\newcommand{\ee}{\end{equation}}

\newcommand{\mati}{\mathcal{I}}
\newcommand{\ud}{{\rm d}}
\newcommand{\ra}{\rightarrow}

\begin{document} 

\mhchemoptions{layout=stacked}

\title{Ultraviolet \ce{H2} luminescence in molecular clouds\\\hspace{0.8cm}induced by cosmic rays}

\titlerunning{Ultraviolet \ce{H2} luminescence in molecular clouds induced by cosmic rays}
\authorrunning{Marco Padovani et al.}

 \author{
          Marco Padovani\inst{1}
          \and
          Daniele Galli\inst{1}
          \and
          Liam H.\ Scarlett\inst{2}
          \and
          Tommaso Grassi\inst{3}
          \and\\
          Una S.\ Rehill\inst{2}
          \and
          Mark C.\ Zammit\inst{4}
          \and
          Igor Bray\inst{2}
          \and
          Dmitry V.\ Fursa\inst{2}
          }

   \institute{$^1$INAF--Osservatorio Astrofisico di Arcetri, 
              Largo E. Fermi 5, 50125 Firenze, Italy\\
              $^2$Department of Physics and Astronomy, 
              Curtin University, Perth, Western Australia 6102, Australia\\
              $^3$Max-Planck-Institut f\"ur extraterrestrische Physik, Giessenbachstrasse 1, 85748 Garching, Germany\\
              $^4$Theoretical Division, Los Alamos National Laboratory, Los Alamos, New Mexico 87545, USA\\
              \email{marco.padovani@inaf.it}
             }

%   \date{}

% \abstract{}{}{}{}{} 
% 5 {} token are mandatory
 
  \abstract
 %context heading (optional)
{Galactic cosmic rays (CRs) play a crucial role in ionisation, dissociation, and excitation processes within dense cloud regions where UV radiation is absorbed by dust grains and gas species. CRs regulate the abundance of ions and radicals, leading to the formation of more and more
complex molecular species, and determine the charge distribution on dust grains.
A quantitative analysis of these effects is essential for understanding the dynamical and chemical evolution of star-forming regions. 
}
{The CR-induced photon flux 
has a significant impact on the evolution of the dense molecular medium in its gas and dust components. This study evaluates the flux of UV photons generated by CRs to calculate the photon-induced dissociation and ionisation rates of a vast number of atomic and molecular species, as well as the integrated UV photon flux.}
{To achieve these goals, we took advantage of recent developments in the
determination of the spectra of secondary electrons, in the calculation of state-resolved excitation cross sections of \ce{H2} by electron impact,
and of photodissociation and photoionisation cross sections.}
{We calculated 
the \ce{H2} level population of each rovibrational level of the $X,B,C,B^\prime,D,B^{\prime\prime},D^\prime$, and $a$ states. We then computed the UV photon spectrum of \ce{H2} in its line and continuum components between 72 and 700~nm, 
with unprecedented accuracy,
as a function of
the CR spectrum incident on a molecular cloud, the \ce{H2} column density, 
the isomeric \ce{H2} 
composition, and the dust properties.
The resulting photodissociation and photoionisation rates are, on average, lower than previous determinations
by a factor of about 2, 
with deviations of up to a factor of 5 for 
the photodissociation of species
such as \ce{AlH}, \ce{C2H2}, \ce{C2H3}, \ce{C3H3}, \ce{LiH}, \ce{N2}, \ce{NaCl}, \ce{NaH}, \ce{O2+}, \ce{S2},
\ce{SiH}, \ce{l-C4}, and \ce{l-C5H}. 
A special focus is given to the photoionisation rates of \ce{H2}, \ce{HF}, and \ce{N2}, as well as to the photodissociation of \ce{H2}, which we find to be orders of magnitude higher than previous estimates.
We give parameterisations for both the photorates and the integrated UV photon flux as a function of the CR 
ionisation rate, which implicitly depends on the \ce{H2} column density, as well as the dust properties.
}
  {}


\maketitle

\section{Introduction}
\label{sec:intro}

The interplay between cosmic-ray (CR) induced processes and the chemical evolution of molecular clouds is a crucial element in our understanding of the astrophysical environment. 
Such findings 
advance our comprehension of fundamental astrophysical phenomena 
and shed light on the interconnections between CR physics and astrochemistry. 
Galactic CRs can penetrate into the densest regions of molecular clouds, where ultraviolet (UV) radiation is blocked by the absorption of dust grains and molecular species. In these dark cloud regions, 
with \ce{H2} column densities higher than about $3$-$4\times10^{21}$~cm$^{-2}$, CRs dominate the processes of ionisation, dissociation, and excitation of gas species. The ionisation of \ce{H2} is the main 
energy loss process of CRs in a molecular cloud, and a key element in the physical and chemical evolution of star-forming regions. The rate of ionisation of \ce{H2} by CRs, $\zeta_{\ce{H2}}$, is a fundamental parameter in non-ideal magnetohydrodynamic simulations and astrochemical models\footnote{In this paper $\zeta_{\ce{H2}}$ is the rate of production of H$_2^+$ ions per \ce{H2} molecule. The dissociative ionisation and double ionisation of \ce{H2}, which produce H$^+$ ions, are not considered here 
\citep[for these processes, see][]{Padovani+2009}.}:

$(i)$ The ionisation fraction, which is a function of $\zeta_{\ce{H2}}$, controls the degree of coupling between the gas and the cloud's magnetic field, influencing the collapse time of molecular clouds.

$(ii)$ Newly formed \ce{H2+} ions react immediately with other hydrogen molecules to form \ce{H3+}, starting a cascade of chemical reactions that lead to the formation of increasingly complex molecules, up to what are believed to be 
the building blocks of terrestrial life \citep{CaselliCeccarelli2012}. 

$(iii)$ The CR-induced dissociation of \ce{H2} 
has important consequences for the formation of complex molecules,
such as \ce{CH3OH} \citep{TielensHagen1982} and \ce{NH3} \citep{Hiraoka+1995,Fedoseev+2015},
on the grain surface through the process of hydrogenation.

$(iv)$ CR-excited rovibrational levels of the ground state of \ce{H2} radiatively decay to the ground level, producing a flux of 
near-infrared (NIR) photons. \citet{Bialy2020} developed a model, extended in \citet{Padovani+2022} and tested observationally in \citet{Bialy+2022} towards dark clouds, showing that from the NIR photon flux it is possible to estimate $\zeta_{\ce{H2}}$ without having to resort to chemical networks and secondary species.

$(v)$ CR-excited electronic states of \ce{H2} radiatively decay to rovibrational levels of the ground state, producing a UV photon flux, mostly in the Lyman-Werner bands, known as the Prasad-Tarafdar effect \citep[][]{PrasadTarafdar1983}. 
These UV photons have a dual effect: they extract electrons from dust grains via the photoelectric effect,
significantly affecting the dust grain charge distribution, and they photodissociate and photoionise atomic and molecular species.

\citet{PrasadTarafdar1983} first presented a quantitative method for estimating the UV emission in the Lyman-Werner bands of \ce{H2} collisionally excited by CR particles. In their pioneering work, they considered a generic excited electronic level (to be interpreted as an appropriate combination of electronic, vibrational, and rotational levels) and a generic excited ground state level other than $v=0$. 
\citet{Sternberg1987} evaluated the Lyman-Werner band emission of CR-excited \ce{H2} and computed the resulting photodissociation rates of several interstellar molecules, focusing in particular on the effects of the CR-generated UV flux on the chemistry of H$_2$O 
and simple hydrocarbons. This work was extended by \citet{Gredel+1987} (specifically to evaluate the CO/C ratio in molecular clouds) and \citet{Gredel+1989}, who included several excited electronic states of \ce{H2} to evaluate photodissociation and photoionisation rates of a large set of molecules. The procedure followed by \citet{Gredel+1989} is summarised in Appendix \ref{app:comparisongredel}.

Recent theoretical and observational developments in the study of the propagation of CRs in molecular clouds, of the production of secondary CR electrons, and the availability of state-resolved cross sections of \ce{H2} excitation by electron impact make it possible to revise and update the results obtained in these studies. In particular, in this work we have taken the following developments into account: 

$(i)$ CR propagation models have shown that the CR ionisation rate 
decreases with \ce{H2} column density,
spanning a large range of orders of magnitude ($10^{-14}-10^{-18}$~s$^{-1}$) for column densities between
$10^{20}$ and $10^{25}$~cm$^{-2}$, depending on the assumptions on the Galactic CR spectrum
\citep[see e.g.][]{Padovani+2009,Padovani+2018a,Padovani+2022}.
This has been supported by a vast number of observational estimates of $\zeta_{\ce{H2}}$ obtained in different environments: 
in diffuse regions of molecular clouds \citep{Shaw+2008,Neufeld+2010,IndrioloMcCall2012,NeufeldWolfire2017,Luo+2023a,Luo+2023b}, 
in low-mass pre-stellar cores \citep{Caselli+1998,MaretBergin2007,Fuente+2016,Redaelli+2021,Bialy+2022}, 
in high-mass star-forming regions \citep{deBoisanger+1996,VanderTak+2000,Hezareh+2008,MoralesOrtiz+2014,Sabatini+2020,Sabatini+2023}, 
in circumstellar discs \citep{Ceccarelli+2004}, 
and in massive hot cores \citep{BargerGarrod2020}.
We refer to Appendix~\ref{app:CRionobstheory} for an updated view of CR ionisation rate estimates from observations
and their comparison with theoretical models.

$(ii)$ Thanks to the methodology recently developed by \citet{Ivlev+2021}, 
it is now possible to rigorously determine the energy spectrum of secondary electrons as a function of the column density for any proton spectrum incident on the cloud 
(see Sect.~\ref{sec:esecspectra} and \citealt{Padovani+2022}),
eliminating the approximation of mono-energetic secondary electrons \citep[usually assumed to have energy of about 30~eV; see e.g.][]{GredelDalgarno1995}.

$(iii)$ Finally, the accuracy of collisional excitation rates of \ce{H2} has substantially increased thanks to the recent availability of molecular convergent close-coupling (MCCC) calculations \citep{Scarlett+2023}, where vibrationally and rotationally resolved electron excitation cross sections of \ce{H2} are computed for a large set of electronic states 
(see Sect.~\ref{sec:xsec}).

As a result of these improvements, we can compute the spectrum of CR-generated UV photons in the wavelength range between 72 and 700~nm with unprecedented accuracy. 
This allows us  
$(i)$ to generate a new set of photodissociation and photoionisation rates of atomic and molecular species 
relevant to astrochemistry and 
$(ii)$ to compute the integrated UV photon flux, namely
the fundamental parameter governing the dust charge distribution
at equilibrium,
as a function of the CR spectrum incident on a molecular cloud, the \ce{H2} column density, 
the isomeric \ce{H2} 
composition, and the dust properties. 
In this work we do not include the UV emission of He and \ce{He+} excited by
electron impact. The calculation of the spectrum of secondary electrons produced by helium ionisation and
the energy loss function of CR protons and electrons propagating 
in helium will be the subject
of a forthcoming study. 
The inclusion of helium may have a significant impact on the ion chemistry in CR-irradiated dark clouds.

The paper is organised as follows: 
In Sect.~\ref{sec:energylevels} we review the physical parameters of the \ce{H2} molecule, which we use in Sect.~\ref{sec:levpop} to compute its rovibrationally resolved level population. 
In Sect.~\ref{sec:UVmethod} we introduce the method for calculating 
the UV emission of \ce{H2} resolved in its rovibrational transitions (lines plus continuum)
presented in Sect.~\ref{sec:UVresults}.
We apply the above results to evaluate 
the photodissociation and photoionisation rates (Sect.~\ref{sec:photorates})
and the integrated UV photon flux (Sect.~\ref{sec:UVflux}),
providing useful parameterisations as a function of the assumption on the interstellar flux of CRs
and the medium composition.
In Sect.~\ref{sec:conclusions} we summarise our main findings.

\section{A model of the H$_2$ molecule: Physical quantities}
\label{sec:energylevels}

This section describes the physical parameters of the \ce{H2} molecule
(energy levels, Einstein coefficients, and collisional excitation cross sections) adopted in this work. 

\subsection{Energy levels}
\label{subsec:energylevels}

Our model for the \ce{H2} molecule includes the following electronic levels: the ground electronic state $1s\sigma$~$^1\Sigma_g^+$ (denoted $X$, 307 rovibrational levels), and the lowest seven electronic excited states that are coupled to the ground state by permitted electronic transitions: the singlet states $2p\sigma$~$^1\Sigma_u^+$ ($B$, 879 levels), $2p\pi$~$^1\Pi_u$ ($C^+$ and $C^-$, 248 and 251 levels, respectively),
$3p\sigma$~$^1\Sigma_u^+$ ($B^\prime$, 108 levels) and  $3p\pi$~$^1\Pi_u$ ($D^+$ and $D^-$, 27 and 336 levels, respectively),
$4p\sigma$~$^1\Sigma_u^+$ ($B^{\prime\prime}$, 160 levels), $4p\pi$~$^1\Pi_u$ ($D^{\prime +}$ and $D^{\prime -}$, 72 and 18 levels, respectively),
and the triplet state $2s\sigma$~$^3\Sigma_g^+$ ($a$, 261 levels). The fully dissociative triplet state $2p\sigma$~$^3\Sigma_u^+$ ($b$) is also included.
Our dataset for the $X$ state is complete, that is to say, radiative and collisional excitation rates are available for all transitions within the ground electronic state. 
For excited electronic states, we considered only those excited rovibrational levels coupled to ground state levels with available radiative and collisional rates. 
The selection rule for rovibrational transitions within the ground electronic state is $\Delta J=0,\pm 2$, between 
$\Sigma$ states is $\Delta J=\pm 1,\pm3,\pm5$, and between $\Sigma$ and $\Pi$ states is $\Delta J=0,\pm 1,\pm2,\ldots,\pm5$. 
In particular, transition to $\Pi$ states $C^-$, $D^-$, and $D^{\prime -}$ have $\Delta J=0,\pm2,\pm4$, while transitions from $\Sigma$ states to $\Pi$ states $C^+$, $D^+$, and $D^{\prime +}$ have $\Delta J=\pm 1,\pm3,\pm5$. Transitions between excited electronic states are not considered. The total number of 1508 rovibrational levels produces 38,970 lines.

Molecular hydrogen occurs in two isomeric forms, para- and ortho-\ce{H2}, depending on the alignment of the two nuclear spins. Given the selection rules above, energy levels of para- (ortho-)\ce{H2} have even (odd) $J$ in the $X$, $C^-$, $D^-$, and $D^{\prime -}$ electronic states,  and odd (even) $J$ in the 
$B$, $C^+$, $B^\prime$, $D^+$, $B^{\prime\prime}$, and $D^{\prime +}$ states. 
Radiative transitions between para- and ortho-\ce{H2} are not allowed.

\subsection{Bound-bound radiative transitions}

Quadrupole (electric and magnetic) and magnetic (dipole) transition probabilities are taken from \citet{Roueff2019} for transitions within the ground electronic state, 
from \citet{Abgrall1993a,Abgrall1993b,Abgrall1993c}  
for transitions between the ground state and the $B$, $C$, $B^\prime$, and $D$ states, and from Glass-Maujean (priv. comm.) for transitions between the ground state and $B^{\prime\prime}$ and $D^\prime$ states. As anticipated in Sect.~\ref{subsec:energylevels}, radiative transitions from $B^{\prime\prime}$ and $D^{\prime +}$ only include the $J=1,\ldots,4$ rotational levels for the $P$ and $R$ branches ($\Delta J=1$ and $\Delta J=-1$, respectively)
and from $D^{\prime -}$ only the $J=1$ level for the $Q$ branch ($\Delta J=0$). In a follow-up paper, we will extend our dataset to levels with higher $J$.

\subsection{Bound-free radiative transitions}

Excited electronic states can decay into the continuum of the ground state, leading to dissociation of \ce{H2}. Radiative transition probabilities to the continuum from the $B$, $C$, $B^\prime$, and $D$ singlet states are taken from by \citet{Abgrall1997,Abgrall2000}\footnote{See also \url{https://molat.obspm.fr}}
and from \citet{Liu+2010}
for transitions from the triplet $a$ state.
There is no available data on the radiative transition probabilities to the continuum from the $B^{\prime\prime}$ and $D^\prime$ states,
but their contribution is expected to be negligible compared to that from the lower Rydberg states (see Sect.~\ref{sec:UVresults}).

We also accounted for the fact that secondary electrons and primary CR protons can yield excited H atoms by direct dissociation and electron capture, resulting in Lyman and Balmer emission with cross sections given by \citet{vanZyl+1989} and \cite{Ajello+1991,Ajello+1996} and by \citet{Mohlmann+1977}, \citet{Karolis+1978}, and \citet{Williams+1982}, respectively.
Radiative transition probabilities for atomic hydrogen are taken from \citet{Kramida+2022}.

\subsection{CR-electron excitation cross sections}
\label{sec:xsec}

Rovibrationally resolved electron-impact cross sections were calculated using the MCCC method \citep{ZFSB17review,Scarlett+2023}.
This is a fully quantum-mechanical method for calculating highly accurate cross sections for electrons and 
positrons scattering on diatomic molecules. \citet{Scarlett+2023} discussed the theoretical and computational 
aspects of the calculation of rovibrationally resolved cross sections for H$_2$, and produced a set of data for all
rotational transitions within the $v=0$ vibrational level of the ground electronic 
state\footnote{Available online at \url{https://mccc-db.org}\label{mccc}}. For the present work,
these calculations have been extended to include all rovibrational transitions 
within the
$X$ state, as well as rovibrationally resolved excitation of the $B$, $C$, $B^\prime$, $D$, $a$, and $b$ states.
For the majority of the transitions, the present MCCC calculations are the first to be performed. For the
pure rotational transitions within the $X$ state, there have been many previous studies, but the MCCC calculations
are the first to incorporate a rigorous account of coupling to the closed inelastic channels, which in all other
calculations was included only approximately via model polarisation potentials \citep[see][for detailed discussion and comparison with previous results]{Scarlett+2023}.

Electron-impact rovibrationally resolved cross sections for $X\rightarrow B^{\prime\prime}$ and $X\rightarrow D^\prime$ are not yet available, but the total cross sections (scattering on $v=0$ only,
summed over all final rovibrational levels) for $B^{\prime\prime}$ is about 0.35 times the $B^\prime$ cross section, and the $D^\prime$ cross section is about 0.4 times
the $D$ cross section~(\cite{Zammit+2017}). Thus, we included these $n=4$ Rydberg states using the above scaling factors. 

\subsection{CR-proton excitation cross sections}

Data on proton-impact excitation of \ce{H2} are limited. Experimental values for the excitation of selected vibrational bands of the Lyman system for proton energies in 
the range 20--130~keV have been reported by \citet{Dahlberg1968}, and for protons above 150~keV by \citet{EdwardsThomas1968}, but the accuracy of these results has been questioned \citep{Thomas1972}. Experimental determinations of the cross sections for the excitation of atomic hydrogen lines of the Balmer series in proton-\ce{H2} collisions have 
been reported by \citet{Thomas1972}, \citet{Williams+1982}, and, more recently, by \citet{DrodzowskiKowalski2018} in the proton energy range 0.2--1.2~keV. It is hoped that these data could be supplied in the near future.  

Due to the scarcity of proton-impact excitation cross sections of electronic states of \ce{H2}, we posit that the proton cross sections are identical to those of electrons of the same velocity, namely
\be
\sigma^{\rm exc, p}(E_p) = \sigma^{\rm exc, e}\left(E_e=\frac{m_e}{m_p}E_p\right)\,,
\ee
where $m_e$ and $m_p$ are the electron and proton mass, respectively, and $E_e$ and $E_p$ their corresponding energies.
With this approximation, protons contribute about 20\% to the total collisional excitation rate at \ce{H2} column densities of the order of
$10^{20}$~cm$^{-2}$, while
above $10^{22}$~cm$^{-2}$ their contribution is negligible ($\lesssim$1\%).

\section{Level populations}
\label{sec:levpop}

In molecular clouds, rovibrational levels of \ce{H2} are populated by radiative excitation due to UV photons of the interstellar radiation field (ISRF), by collisions with CR particles (mostly protons, primary and secondary electrons), and by collisions with ambient particles; \ce{H2} levels are depopulated by spontaneous emission and de-excitation by collision with ambient particles.

The equations governing the population and depopulation of a level 
$XvJ$ the ground electronic level and of a generic excited electronic level $SvJ$ are
\begin{eqnarray}
\label{popX}
n_{XvJ}\left(\sum_{v^\prime\!J^\prime < vJ}A_{XvJ\ra X v^\prime\!J^\prime} + 
\sum_{v^\prime\!J^\prime} C_{XvJ\ra X v^\prime\!J^\prime}+
\sum_{S v^\prime\!J^\prime} C_{XvJ\ra S v^\prime\!J^\prime}\right) = 
\nonumber \\
\sum_{v^\prime\!J^\prime > vJ}n_{X v^\prime\!J^\prime}A_{X v^\prime\!J^\prime \ra XvJ}
+\sum_{v^\prime\!J^\prime}n_{X v^\prime\!J^\prime}C_{X v^\prime\!J^\prime\ra XvJ}
+\sum_{S v^\prime\!J^\prime}n_{S v^\prime\!J^\prime} A_{S v^\prime\!J^\prime\ra XvJ}
\end{eqnarray}
and
\begin{eqnarray}
\label{popS}
n_{S v J}\left(\sum_{v^\prime\!J^\prime < vJ} A_{SvJ\ra X v^\prime\!J^\prime}
+ \int \tilde{A}_{S v J \ra Xc} \,\ud \nu\right)=\sum_{v^\prime\!J^\prime}n_{X v^\prime\!J^\prime} C_{X v^\prime\!J^\prime \ra SvJ}\,,
\end{eqnarray}
respectively. 
In the above equations, the transitions' vibrational and rotational quantum numbers are labelled $v,v^\prime$, and $J,J^\prime$, respectively,
while the subscript $c$ represents the continuum of the ground electronic state. 
Rates of spontaneous emission are denoted by $A_{kk^\prime}$, while excitation and de-excitation 
rates by $C_{kk^\prime}$, 
where $kk^\prime$ denotes a generic transition from a level $k$ to a level $k^\prime$ in shorthand notation (i.e. $k$ represents a triplet $XvJ$ or $SvJ$). 
The excitation and de-excitation rates, $C_{kk^\prime}$, contain a contribution from photons of 
the ISRF ($C_{kk^\prime}^{\rm rad}$),
and a contribution from collisions with CRs and cloud particles. Stimulated emission is neglected. Collisional excitation is dominated by CR electrons ($C_{kk^\prime}^{\rm CR}$) and collisional de-excitation by cloud particles ($C_{kk^\prime}^{\rm cl}$), in particular by collisions with ambient \ce{H2}. Thus, $C_{kk^\prime}=C_{kk^\prime}^{\rm rad}+C_{kk^\prime}^{\rm CR}+C_{kk^\prime}^{\rm cl}$.
 
As shown by Eq.~(\ref{popS}), \ce{H2} in rovibrational levels within an excited electronic state, $S,$ can either decay back to a bound rovibrational level of the ground electronic 
state, $X,$ with transition probability $A_{SvJ\ra X v^\prime\!J^\prime}$ or decay into the 
continuum of the ground state and dissociate, with the transition probability per unit frequency $\tilde{A}_{SvJ\ra Xc}$. In the latter case, we assume that the dissociated \ce{H2} converts back to \ce{H2} in the $v=0$ level of the ground state 
and is partitioned into the $J=0$ and $J=1$ rotational levels depending on the assumed \ce{H2} ortho-to-para (o:p) ratio.

We now turn to examine the excitation processes in detail.

\subsection{Radiative excitation}
\label{sec:esecspectra}

Radiative excitation rates are
\be
C^{\rm rad}_{kk^\prime}(N)=B_{kk^\prime}J_\nu^{\rm ISRF}(N)\,,
\ee
where 
\be
B_{kk^\prime}=\left(\frac{g_{k^\prime}}{g_k}\right)\frac{c^2}{2h\nu_{kk^\prime}^3} A_{k^\prime\! k}\,,
\ee
$g_{k}$ ($g_{k^\prime}$) is the degeneracy of state $k$ ($k^\prime$),
and 
$J_\nu^{\rm ISRF}(N)$ is the mean intensity of the 
ISRF at the transition frequency $\nu_{kk^\prime}$ and at the \ce{H2} column density $N$. 
We adopted the radiation field of \citet{Draine1978} and \citet{vanDishoeckBlack1982}, whose intensity is characterised by a scaling factor $\chi$ ($\chi=1$ corresponds to 
the UV ISRF field at 100~nm).

\subsection{CR collisional excitation}
\label{colexcrate}

To obtain CR excitation rates, the cross sections described above must be folded with the spectrum of CR particles (protons, primary and secondary electrons)\footnote{The particle spectrum is defined as the number of particles per unit energy, time, area, and solid angle.}. For each transition $k\ra k^\prime$, CR-induced collisional excitation rates (dropping 
the subscript $kk^\prime$ for simplicity) are
\be
C^{\rm CR}(N) = \sum_i\Omega_s\int j_i(E,N)\sigma_{s}^{\rm exc}(E)\,\ud E\,,
\label{gammacoll}
\ee
where $i$ is the CR particle species considered (CR protons, primary CR electrons, and secondary electrons),
$j_s(E,N)$ is the particle spectrum at column density $N$ (see Sect.~\ref{sec:esecspectra}) and
$\sigma^{\rm exc}_s(E)$ is the excitation cross section.
Taking into consideration a semi-infinite slab configuration, a value of $\Omega_s=2\pi$~sr is assigned to primary CR nuclei and electrons, while secondary electrons, 
generated locally and exhibiting nearly isotropic propagation, are attributed a value of $\Omega_s=4\pi$~sr.

\subsubsection{Spectrum of secondary electrons}
\label{sec:CRspectrum}

For the calculation of the CR spectrum of secondary electrons 
at depth $N$ into a semi-infinite cloud, we followed the modelisation 
developed by \citet{Padovani+2009} and \citet{Ivlev+2021}. 
In particular, at the cloud's surface 
we assumed an interstellar CR spectrum parametrised as in \citet{Padovani+2018a},
\be\label{CRfit}
j_i^{\rm IS}(E) = C\frac{E^\alpha}{(E+E_0)^\beta}~{\rm eV^{-1}~s^{-1}~cm^{-2}~sr^{-1}}\,,
\ee
where $i=e,p$ and $E$ is in eV. 
Primary CR electrons have a negligible effect on both 
ionisation and excitation of \ce{H2} 
above column densities of $\approx10^{21}$~cm$^{-2}$ \citep[see][]{Padovani+2022}. Therefore, we considered a single spectrum of CR electrons. 
For primary CR protons, we explored values of the low-energy spectral slope $\alpha$ ranging from $\alpha=0.1$ to $\alpha=-1.2$. 
In particular, we show results for $\alpha=0.1$ (labelled 'low' spectrum, $\mathscr{L}$), which reproduces the proton flux detected by the Voyager probes;
$\alpha=-0.8$ (labelled 'high' spectrum, $\mathscr{H}$), which results in an average value of the ionisation rate estimated in diffuse molecular regions; 
and $\alpha=-1.2$ (labelled 'upmost' spectrum, $\mathscr{U}$), which produces values of the ionisation rate that match the upper envelope of the available observational estimates 
(see \citealt{Padovani+2022} and Appendix~\ref{app:CRionobstheory} for an updated plot of the CR ionisation measurements).
Table~\ref{tab:jis} lists the parameters for the Galactic CR spectrum.

\begin{table}[!h]
\caption{Parameters of the Galactic CR electron and proton spectra, Eq.~(\ref{CRfit}).}
\begin{center}
\begin{tabular}{lcccc}
\toprule\toprule
CR species ($i$) & $C$ & $E_{0}/\rm{eV}$ & $\alpha$ & $\beta-\alpha$\\
\midrule
$e$ & $2.1\times10^{18}$ & $7.1\times10^8$ & $-1.3$ & 3.2\\
$p$ (model $\mathscr{L}$) & $2.4\times10^{15}$ & $6.5\times10^8$ & $\phantom{-}$0.1 & 2.7\\
$p$ (model $\mathscr{H}$) & $2.4\times10^{15}$ & $6.5\times10^8$ & $-0.8$ & 2.7\\
$p$ (model $\mathscr{U}$) & $2.4\times10^{15}$ & $6.5\times10^8$ & $-1.2$ & 2.7\\
\bottomrule
\end{tabular}
\tablefoot{$C$ is in units of ${\rm eV^{-1}\ s^{-1}\ cm^{-2}\ sr^{-1}}$.}
\end{center}
\label{tab:jis}
\end{table}%

The interstellar spectrum given by Eq.~(\ref{CRfit}) 
is propagated deep into the cloud under the assumption of
continuous slowing-down approximation \citep[see e.g.][]{Takayanagi1973} adopting
the recently updated energy loss functions of protons and electrons colliding with \ce{H2} (see Appendix~\ref{app:lossfunctions}). 
Figure~\ref{fig:esec-spectra} shows the secondary-electron spectra at $10^{20}$ and $10^{24}$~cm$^{-2}$ 
derived from the three CR proton models ($\mathscr{L,H,U}$)
compared to some representative electron excitation cross sections.
The largest contribution to $C^{\rm CR}$ 
occurs around $E \approx 20$~eV for excitation to the singlet $B,C,B'$, and $D$ states and $E \approx 14$~eV for excitation to the triplet $a$ state.
However, to recover 95\% of the CR excitation rate for the singlet $B,C,B'$, and $D$ states, it is necessary to integrate the cross sections from the threshold energy up to $\sim 250$~eV  
(but only and up to $\sim 20$~eV 
for the triplet $a$ state, given the sharp decline in energy of the cross section).
%The purpose of this 95\% fiducial level is to demonstrate that 
Thus, the majority of collision-induced excitation to singlet electronic states stems from
a broad energy range of secondary electrons. 
In contrast, considering 30~eV monoenergetic electrons, only about 30\% of the total collisional excitation rate is recovered, reinforcing
the importance of considering the secondary electron spectrum in its entirety.

\subsection{De-excitation by collisions with ambient cloud H$_2$}

We included in the coefficients $C_{kk^\prime}^{\rm cl}$ the 
de-excitation of rotational transitions in the $v=0$ level of the ground electronic state of \ce{H2} 
due to collisions with ambient cloud \ce{H2} \citep{Hernandez+2021}. We ignored 
collisions with He and other species. 

\begin{figure}
\includegraphics[width=0.5\textwidth]{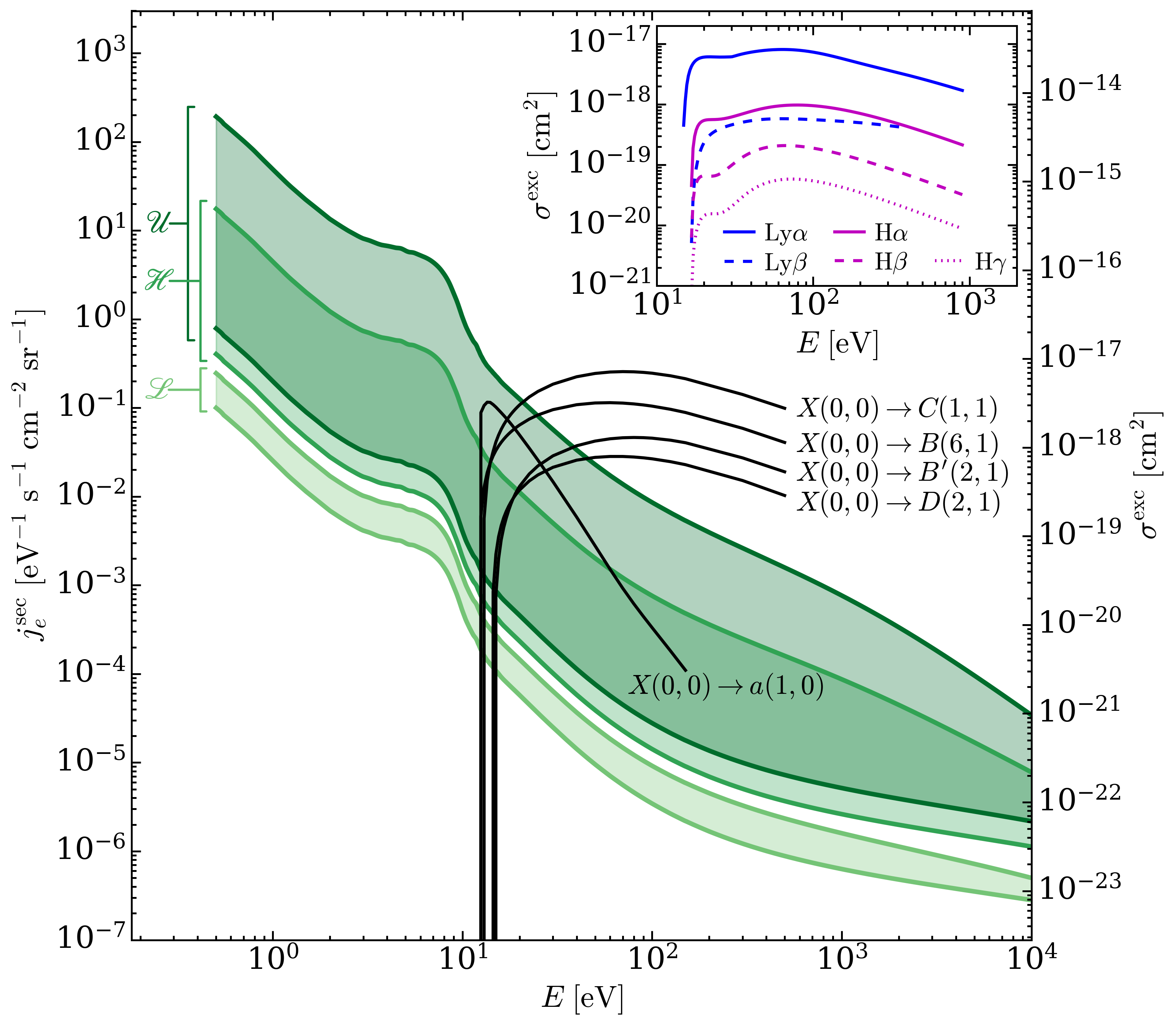}
\caption{Secondary-electron spectra compared to some electron excitation cross sections for singlet $B$, $C$, $B^\prime$, and $D$ and triplet $a$ states as a function of the energy \citep{Scarlett+2023}.
The green-scale shaded regions enclose the secondary-electron spectra calculated at $N=10^{20}$\;cm$^{-2}$ (upper boundary) and $10^{24}$\;cm$^{-2}$ (lower boundary) 
for the three Galactic CR proton models ($\mathscr{L,H,U}$). 
The five cross sections shown in black solid lines are the strongest ones for the excitation of the $B$, $C$, $B^\prime$, and $D$ singlet states and the triplet $a$ state
from the ground state $X(0,0)$. The inset shows the electron excitation cross section of atomic hydrogen leading to Lyman and Balmer lines \citep{Mohlmann+1977,Karolis+1978,vanZyl+1989,Ajello+1991,Ajello+1996}.}
\label{fig:esec-spectra}
\end{figure}

\subsection{Comparison between radiative and CR excitation rates}
\label{sec:colradexcrates}

\begin{figure}
\includegraphics[width=0.5\textwidth]{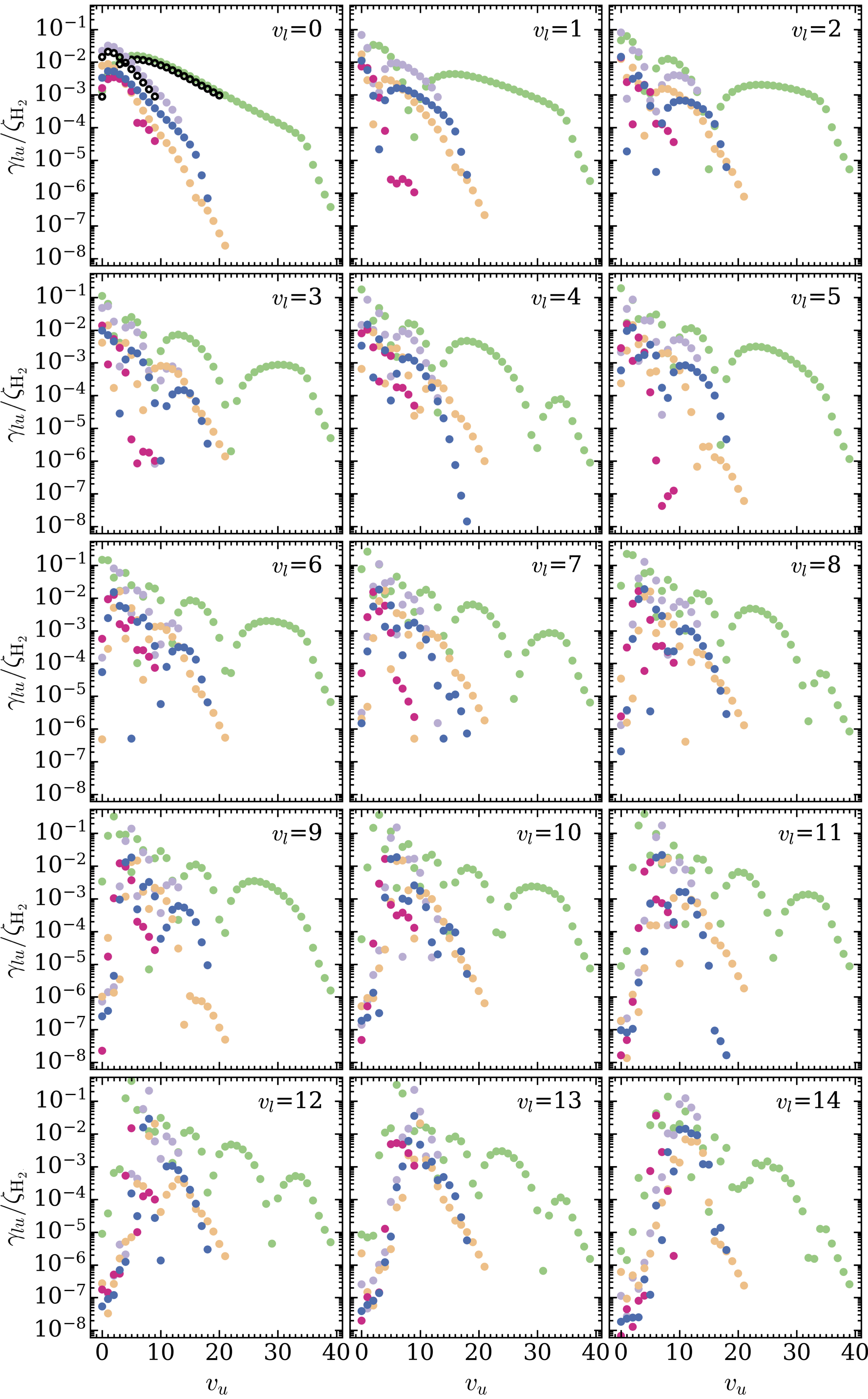}
\caption{Ratio between the $e$-$\ce{H2}$ excitation rates summed over the initial and final
rotational states and the CR ionisation rate for 
model $\mathscr{H}$ and $N=10^{20}$~cm$^{-2}$ as a function of the upper vibrational level, $v_u$. Each
subplot shows results for a different lower vibrational level, $v_l$.
Solid green, violet, magenta, blue, and orange circles refer to the excitation of the $B$, $C$, $B^\prime$, $D$, and $a$ states from the ground
state $X$, respectively. Empty black circles show a subset of previous estimates by \citet{Cecchi-PestelliniAiello1992} (see Sect.~\ref{sec:colradexcrates} for more details).}
\label{collisionratesCP92electrons}
\end{figure}

Figure~\ref{collisionratesCP92electrons} shows 
our new set of rovibrationally resolved collisional excitation rates normalised to 
the CR ionisation rate, $\zeta_{\ce{H2}}$, obtained
by considering the three Galactic CR proton spectra $\mathscr{L}$, $\mathscr{H}$, and $\mathscr{U}$ introduced in Sect.~\ref{sec:CRspectrum}.
In this figure we only show the results for model $\mathscr{H}$ and $N=10^{20}$~cm$^{-2}$ since the
ratio $C^{\rm CR}/\zeta_{\ce{H2}}$ is almost independent of the column density,
decreasing by only $\sim$20\% from $10^{20}$ to $10^{24}$~cm$^{-2}$.
Collisional excitation rates increase 
when using CR proton spectra with a larger component at low energies,
since secondary electron fluxes are larger as well (see Fig.~\ref{fig:esec-spectra}),
and are on average in the following ratios:
$C^\mathscr{U} : C^\mathscr{H} : C^\mathscr{L} \simeq 90:9:1$ for the $X\ra B,C,B^\prime,D$
transitions and $80:8:1$ for the $X\ra a$ transitions at $N=10^{20}$~cm$^{-2}$ with a dispersion of $\approx$2\%.
These ratios tend to be smaller at larger column densities, since low-energy CRs are stopped locally, and the
local CR spectrum becomes independent of the assumption on the low-energy slope. For example, at $N=10^{24}$~cm$^{-2}$
the above ratios become $C^\mathscr{U} : C^\mathscr{H} : C^\mathscr{L} \simeq 5:3:1$ for
$X \ra B,C,B^\prime,D,a$ transitions with a dispersion of $\lesssim$1\%.

The top-leftmost 
panel of Fig.~\ref{collisionratesCP92electrons} shows the comparison of our $C^{\rm CR}/\zeta_{\ce{H2}}$
ratios with those
previously computed by \citet{Cecchi-PestelliniAiello1992}. The latter are 
limited to the fundamental vibrational level of the ground state, $v=0$, and use the cross 
sections for singlet-state excitation by \citet{Shemansky+1985}. Since these cross sections are not rotationally 
resolved, in this figure we sum our collisional excitation rates over all
the initial and final rotationally states. 
Besides, we limit the comparison up to $v=20$ for $X\ra B$ and $v=9$ for $X\ra C$ for a better comparison with \citet{Cecchi-PestelliniAiello1992}.
Our new collisional excitation rates are between 30 and 60\% higher than those of \citet{Cecchi-PestelliniAiello1992}, depending on the CR proton spectrum assumed and the \ce{H2} column density.

\begin{figure}
\begin{center}
\includegraphics[width=0.5\textwidth]{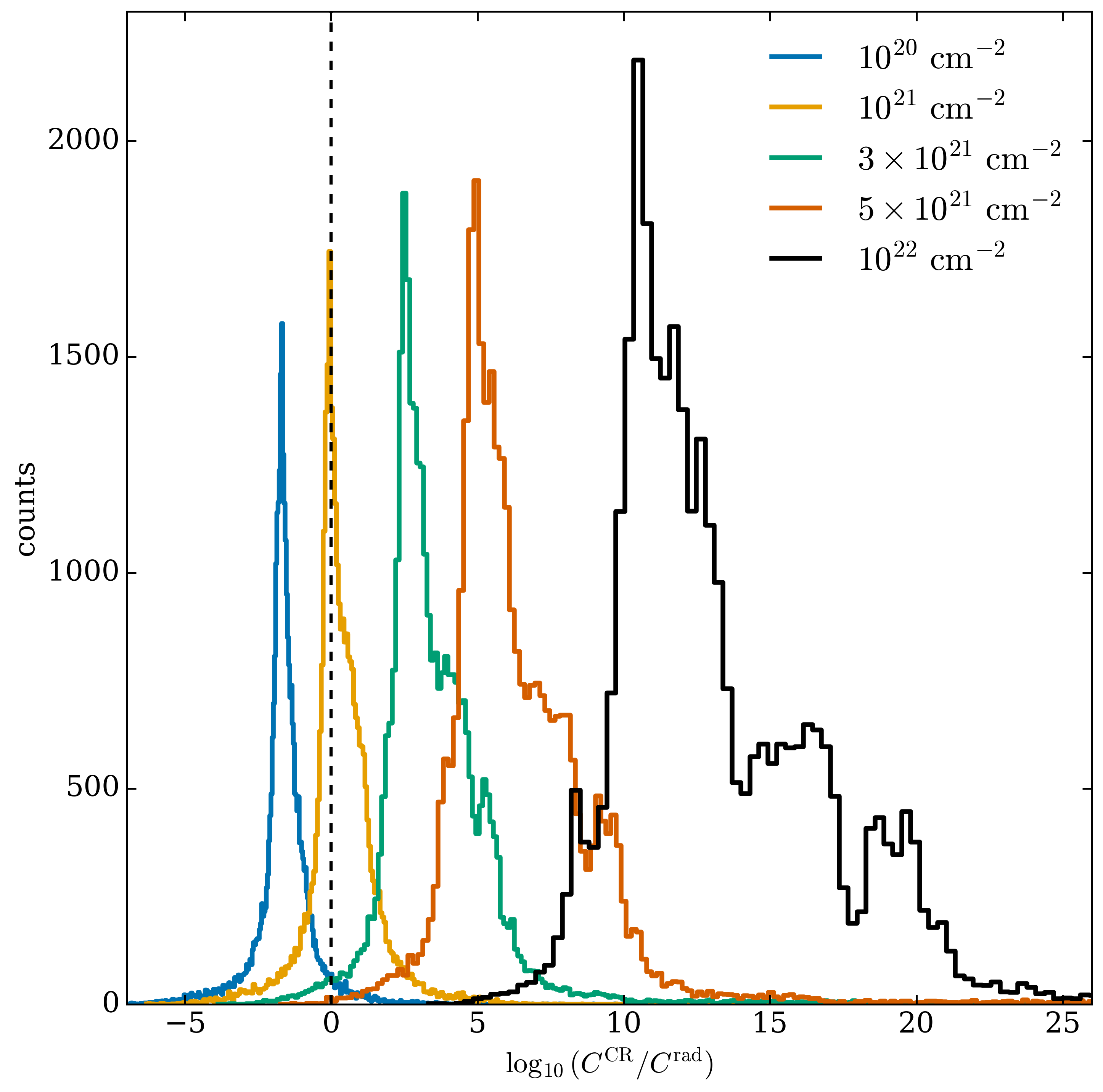}
\caption{Histograms of the ratio between collisional and radiative excitation rates, $C^{\rm CR}/C^{\rm rad}$, for five
representative column densities, from $10^{20}$ to $10^{22}$~cm$^{-2}$, for
an interstellar UV field with $\chi=1$, $R_V=3.1,$ and model $\mathscr{H}$.
The black dashed line shows where collisional and radiative processes
equally contribute to excitation of \ce{H2}.}
\label{compare_radexc_vs_collexc}
\end{center}
\end{figure}

We can compare the relative 
contribution of radiative and collisional excitation rates as a function of column density.
Figure~\ref{compare_radexc_vs_collexc} shows the histograms of the logarithmic ratio 
$\log_{10}(C^{\rm CR}/C^{\rm rad})$ for all $(X),v,J \ra (B,C,B^\prime,D),v^\prime,J^\prime$ common transitions
of the two excitation mechanisms as a function of $N$, for an interstellar UV field with 
$\chi=1$, and for model $\mathscr{H}$.
We also accounted for different values of $R_V$, which is 
a measure of the slope of the extinction at visible wavelengths,
computed by \citet{LiDraine2001}, \citet{WeingartnerDraine2001}, and \citet{Draine2003a,Draine2003b,Draine2003c} 
for different mixed grain sizes and composition\footnote{\url{www.astro.princeton.edu/~draine/dust/dustmix.html}}.

In Fig.~\ref{compare_radexc_vs_collexc} we consider $R_V=3.1$.
At column densities lower than $10^{21}$~cm$^{-2}$,
radiative excitation dominates since interstellar UV photons are 
still able to penetrate molecular clouds. At $N\approx 10^{21}$~cm$^{-2}$
the two excitation processes contribute on average with equal weight. At larger column densities, the interstellar UV field 
is completely attenuated, and CRs, in particular secondary electrons, remain the only agents controlling the populations of the excited levels of \ce{H2}.
As $\chi$ varies, the ratio $C^{\rm CR}/C^{\rm rad}$ changes by the
inverse of the same factor.

Although in this paper we mainly show the results for $R_V=3.1$, we also considered the cases 
$R_V=4.0$ and $5.5$. As $R_V$ increases, the distribution of $\log_{10}(C^{\rm CR}/C^{\rm rad})$ 
is shifted to smaller values as $N$, meaning that radiative excitation dominates
deeper into the cloud.
This is because, as $R_V$ increases, the extinction cross section decreases in the 
wavelength range $72-180$~nm, namely where \ce{H2} bound-bound transitions occur (see Sect.~\ref{sec:UVresults}).

\begin{figure}
\includegraphics[width=0.5\textwidth]{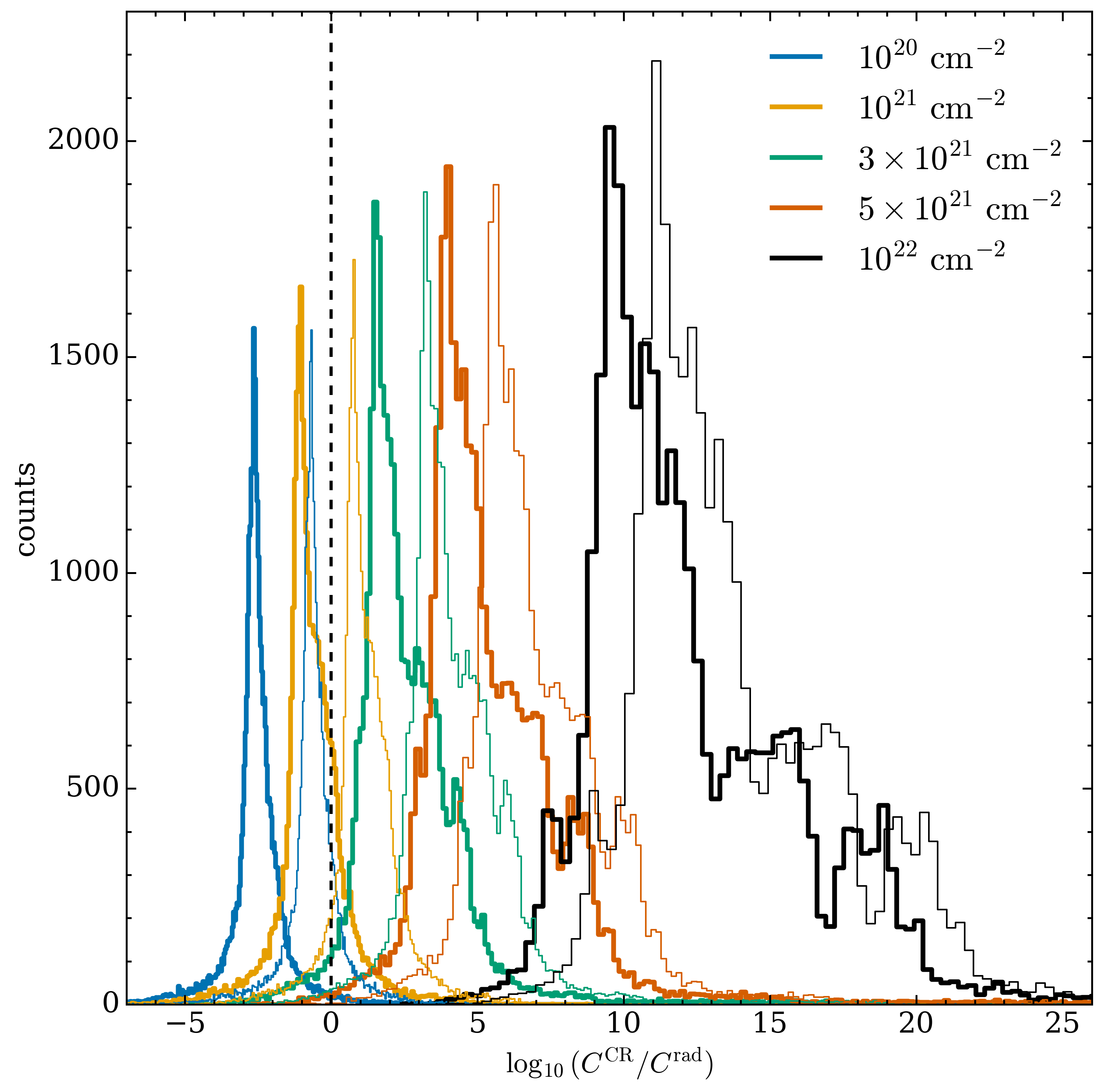}
\caption{Same as Fig.~\ref{compare_radexc_vs_collexc}, but for models $\mathscr{L}$ (thick lines) and $\mathscr{U}$ (thin lines).}
\label{fig:compare_radexc_vs_collexc_Landm12}
\end{figure}

For the sake of completeness, 
in Fig.~\ref{fig:compare_radexc_vs_collexc_Landm12} we show histograms of the logarithmic ratio $\log_{10} (C^{\rm CR}/C^{\rm rad})$ for models $\mathscr{L}$ and $\mathscr{U}$.
As might be expected, since model $\mathscr{L}$ has a smaller flux of CR protons at low energies,  the histogram has the same pattern as that for the model $\mathscr{H}$, %(Fig.~\ref{compare_radexc_vs_collexc}), 
but shifted to smaller $\log_{10} (C^{\rm CR}/C^{\rm rad})$ values. 
This means that higher column densities are required for collisional excitation to dominate radiative excitation. 
The opposite is true for model $\mathscr{U}$: due to the higher flux of CR protons at low energies, collisional excitation already dominates radiative excitation at lower column densities.

A comparison with the $C^{\rm CR}$ computed
by \citet{Gredel+1989} is presented in Appendix~\ref{app:comparisongredel}.

\section{Ultraviolet photon spectrum: Methodology}
\label{sec:UVmethod}

\begin{figure}
\includegraphics[width=0.5\textwidth]{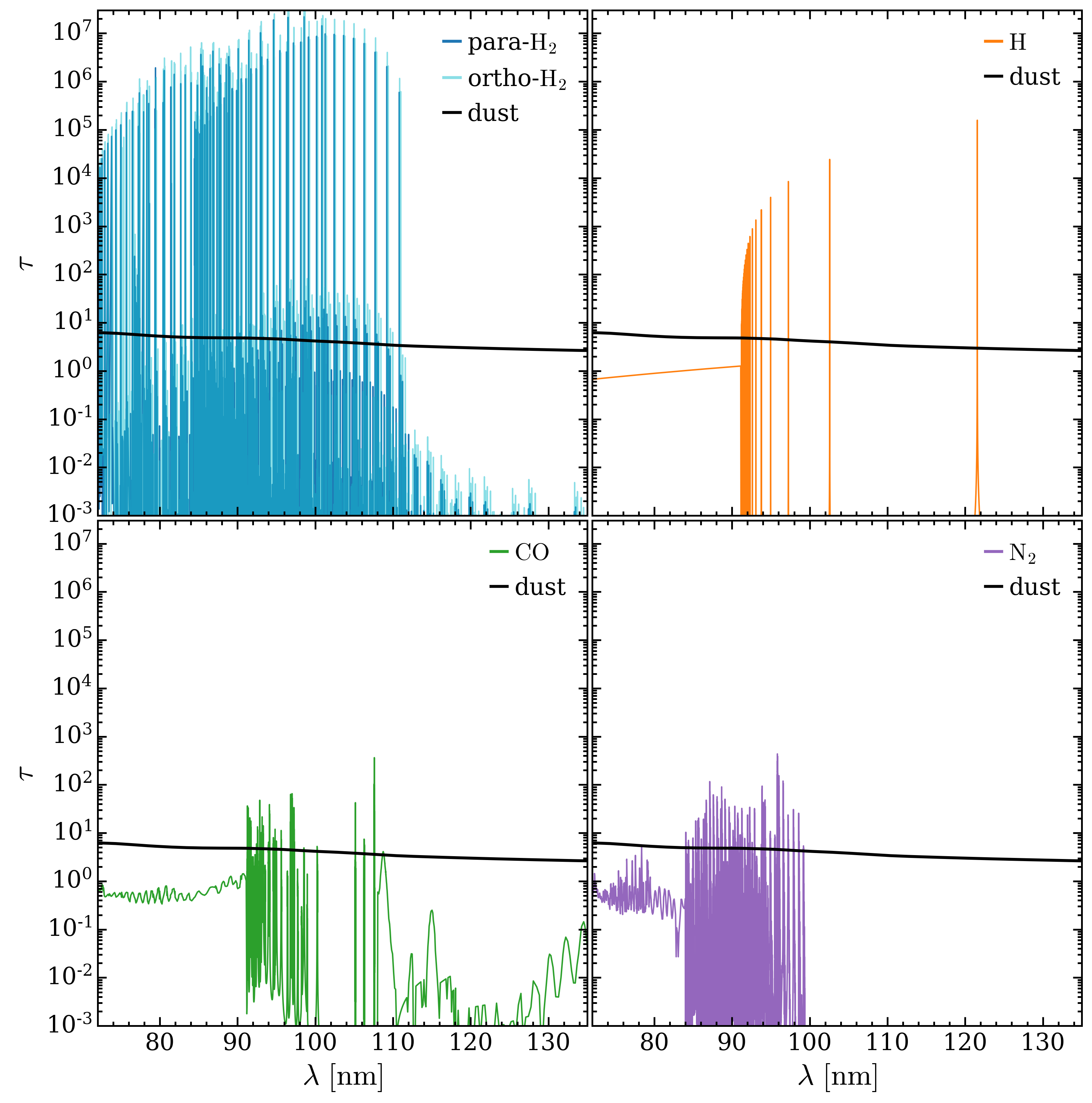}
\caption{Comparison of the optical depth, $\tau$, of dust (black solid lines) to that of
para- and ortho-\ce{H2} (blue and cyan solid lines, respectively, upper-left panel),
H (orange solid lines, upper-right panel),
CO (green solid lines, bottom-left panel),
and \ce{N2} (purple solid lines, bottom-right panel)
at $N(\ce{H2})=10^{21}$~cm$^{-2}$.
The \ce{H2} optical depth is shown for model $\mathscr{H}$ 
and \ce{H2} o:p=3:1.
The H, CO, and \ce{N2} optical depths have been computed by assuming
constant abundances ($x_{\rm H}=10^{-4}$, $x_{\rm CO}=x_{\rm N_2}=10^{-5}$)
and the photoabsorption cross sections presented in the
Leiden Observatory database$^{\ref{fnote:photoratedatabase}}$
\citep[see also][]{Heays+2017}.}
\label{fig:opacity}
\end{figure}

Thanks to the availability of rotationally resolved spontaneous emission 
coefficients and collisional excitation rates, 
we were able to determine 
the level population of each rovibrational level of the $X,B,C,B^\prime,D,B^{\prime\prime},D^\prime$ and $a$ states
(Eqs.~\ref{popX} and~\ref{popS}) and calculate the UV emission of \ce{H2} resolved in its rovibrational transitions
(lines plus continuum), as a function of the assumed interstellar CR spectrum, 
\ce{H2} column density $N$, intensity of the interstellar UV field $\chi$, \ce{H2} o:p ratio, and dust properties.
To compute the radiative transfer of UV radiation in a semi-infinite plane-parallel cloud with embedded sources,
we adopted the spherical harmonics method of  \citet{Roberge1983}.
The specific intensity of radiation in the cloud, $I_\nu(N,\mu)$, is expanded in a truncated series of finite odd order $L$ of Legendre polynomials, $P_\ell(\mu)$, as
\be\label{Inu}
I_\nu(N,\mu) = \sum_{\ell=0}^L (2\ell+1) f_{\nu,\ell}(N) P_\ell(\mu)\,,
\ee
where $\mu$ is the cosine of the angle of propagation of the radiation.
The mean intensity, 
$J_\nu(N)$, in units of energy per unit area, time, frequency, and solid angle, is defined as the average of the specific intensity over all solid angles, that is,
\be\label{eq:Jnu}
J_\nu(N) = \frac{1}{4\pi}\int I_\nu(N,\mu)\ud\Omega = f_{\nu,0}(N)\,, 
\ee
where $f_{\nu,0}(N)$ is the first coefficient ($\ell=0$) in Eq.~(\ref{Inu}).
The specific flux, $F_\nu(N)$, in units of energy per unit area, time, and frequency, is
\be\label{eq:Fnu}
F_\nu(N) = \int I_\nu(N,\mu)\mu\ud\Omega\,
= 4\pi f_{\nu,1}(N)\,,
\ee
so that the integrated flux, $\Phi_{\rm UV}$, in units of photons per unit area and time, is
\be\label{eq:Nph}
\Phi_{\rm UV}(N)=\int \frac{F_\nu(N)}{h\nu}\ud\nu\,,
\ee
where $h$ is the Planck constant.
Following \citet{Roberge1983}, $f_{\nu,\ell}(N)$ at a given column density can be expressed as the sum of two terms, $f^-_{\nu,\ell}(N)$ and $f^+_{\nu,\ell}(N)$, representing the contribution 
from 0 to $N$ and from $N$ to $N_{\rm tot}$, respectively, where $N_{\rm tot}$ is the total observed line-of-sight \ce{H2} column density.
These two terms read as 
\begin{eqnarray}\label{f0minus}
f^-_{\nu,\ell}(N) &=& \sum_{m=1}^M R_{\ell,m}\Biggl\{C_{-m}\exp\Biggl[k_m\Biggl(2\sigma_\nu^{\rm att}N +\Biggr.\Biggr.\Biggr.\\\nonumber
&&\Biggl.\Biggl.\Biggl.\sum_{kk^{\prime}}\int_0^N \sigma_{\nu,kk^\prime} x_k(N^\prime)\ud N^\prime\Biggr)\Biggr] + \hat{Z}_{\nu,m}(N)\Biggr\}
\end{eqnarray}
and
\begin{eqnarray}\label{f0plus}
f^+_{\nu,\ell}(N) &=& \sum_{m=1}^M R_{\ell,m}\Biggl(C_{m}\exp\Biggl\{-k_m\Biggl[2\sigma_\nu^{\rm att}\left(N_{\rm tot}-N\right) +\Biggr.\Biggr.\Biggr.\\\nonumber
&&\Biggl.\Biggl.\Biggl.\sum_{kk^{\prime}}\int_N^{N_{\rm tot}} \sigma_{\nu,kk^\prime} x_k(N^\prime)\ud N^\prime\Biggr]\Biggr\} + \hat{W}_{\nu,m}(N)\Biggr)\,.
\end{eqnarray}
Here, $m=\pm1,\ldots,\pm M$, where $M=(L+1)/2$, $k_m$ and $R_{\ell,m}$
are 
the eigenvalues and the elements of the eigenvector matrix, respectively, of the associated system of linear
first-order differential equations (see Appendix of \citealt{Roberge1983} for details); $\sigma_{\nu,kk^\prime}$ are the \ce{H2} photoabsorption cross sections
from level $k$, with fractional abundance $x_k$, to level $k^\prime$
and the sum over $kk^{\prime}$ in Eqs.~(\ref{f0minus}) and~(\ref{f0plus}) extends over 
all transitions that contribute to absorption at frequency $\nu$; 
$\sigma_\nu^{\rm att}=\sigma_\nu^{\rm ext}+\sum_s x_s\sigma_{\nu,s}^{\rm abs}$ is the total attenuation cross section per hydrogen nucleus
given by the sum of the dust extinction cross section per hydrogen nucleus ($\sigma_\nu^{\rm ext}$), which is a function
of $R_V$ \citep{Draine2003a}, and the photoabsorption cross sections of gas species $s$ with abundance relative to hydrogen nuclei equal to $x_s$.
We only included the most abundant species in the gas phase (H, CO, and \ce{N2}) that can cause absorption (shielding) of the photon flux
at the \ce{H2} column densities of interest, namely above $3$-$4\times10^{21}$~cm$^{-2}$, where the UV ISRF is completely attenuated. 
For simplicity, following \citet{Heays+2017}, we assumed constant abundances for these three species\footnote{The case of non-constant abundances can be easily accounted for. 
\citet{Padovani+2018b} show the profile of $x_{\rm H}$ for the interstellar CR proton spectra considered in this work.}: $x_{\rm H}=10^{-4}$ and $x_{\rm CO}=x_{\rm N_2}=10^{-5}$.
The photoabsorption cross sections $\sigma_{\nu,s}^{\rm abs}$ are taken 
from the Leiden Observatory database\footnote{\url{http://www.strw.leidenuniv.nl/~ewine/photo}\label{fnote:photoratedatabase}} \citep[see also][]{Heays+2017}.
For each transition $k\ra k^\prime$, the \ce{H2} photoabsorption cross section is
\be
\sigma_{\nu,kk^\prime}=\frac{1}{4\pi}B_{kk^\prime} h\nu_{kk^\prime}\phi_\nu\,,
\ee
where $\phi_\nu$ is the line profile,  
and $\nu$ is the frequency of the transition.
We note that we were able to treat \ce{H2} self-absorption line by line 
since with the procedure described in Sect.~\ref{sec:energylevels} we could calculate the population of each rovibrational level.
Figure~\ref{fig:opacity} shows the contribution of dust and gas (\ce{H2}, H, CO, and \ce{N2}) 
absorption
to the optical depth at the \ce{H2} column density $N=10^{21}$~cm$^{-2}$. 
We note that the overlapping line wings
of the \ce{H2} transitions plus the CO, \ce{N2}, and H lines produce a net opacity 
that exceeds the dust opacity over a large fraction of the wavelength range 72$-$130~nm.
Therefore, it is important to include in Eqs.~(\ref{f0minus}) and (\ref{f0plus}) an explicit sum over all
transitions contributing to the emission at frequency $\nu$ in order to account
for overlaps and blending.

The functions $\hat{Z}_{\nu,m}(N)$ and $\hat{W}_{\nu,m}(N)$ describe the contribution of isotropic embedded sources 
and are
\be
\hat{Z}_{\nu,m}(N) = \frac{k_m}{1-\omega_\nu}R_{m,0}^{-1}\mati_{\nu,m}^Z(N)
\ee
and
\be
\hat{W}_{\nu,m}(N) = \frac{k_m}{1-\omega_\nu}R_{m,0}^{-1}\mati_{\nu,m}^W(N)\,,
\ee
where $\omega_\nu$ is the dust albedo, 
while
\begin{eqnarray}
\mati_{\nu,m}^Z(N)&=&\int_0^N S_\nu(N')%
\exp\left[-2\sigma_\nu^{\rm att}k_m(N-N')\right]\times\\\nonumber
&&\exp\left[-k_m\sum_{kk^{\prime}} \int_{N'}^N \sigma_{\nu,kk^\prime} x_k(N'') \ud N''\right] \ud N'
\end{eqnarray}
and
\begin{eqnarray}
\mati_{\nu,m}^W(N)&=&\int_N^{N_{\rm tot}} S_\nu(N')%
\exp\left[-2\sigma_\nu^{\rm att}k_m(N'-N)\right]\times\\\nonumber
&&\exp\left[-k_m\sum_{kk^{\prime}} \int_N^{N'} \sigma_{\nu,kk^\prime} x_k(N'') \ud N''\right] \ud N'\,.
\end{eqnarray}
Here, the source function is
\be\label{eq:S0}
S_\nu(N) = \frac{1}{4\pi}\left[\sum_{kk^{\prime}} x_{k^\prime}(N) A_{k^\prime\! k}h\nu_{kk^\prime} \phi_\nu + 
\sum_{k^{\prime}}x_{k^\prime}(N) \tilde{A}_{k^\prime c} E_t \right]\,,
\ee
where $A_{k^\prime\! k}$ is the transition probability from 
level $k^\prime$ to $k$, and $\tilde{A}_{k^\prime c}$
is the transition probability per unit frequency from level $k^\prime$ to the 
continuum of the ground electronic state ($c$), with transition energy $E_t$.

Finally, the constants $C_m$ in Eqs.~(\ref{f0minus}) and~(\ref{f0plus}) are obtained from Mark's conditions \citep{Mark1944,Mark1945}
by inverting the system of equations
\be
\sum_{m=-M}^M \mathcal{B}_{im} C_m = Q_i,
\ee
with $i=1,\ldots,2M$, where
{\footnotesize{
\be
  \mathcal{B}_{im} =
    \begin{cases}
      \sum_{\ell=0}^L (2\ell+1)P_\ell(\mu_i)R_{\ell m}\,, & \text{$(\mu_i<0 \land m<0) \lor (\mu_i>0 \land m>0)$}\\
      \sum_{\ell=0}^L (2\ell+1)P_\ell(\mu_i)R_{\ell m}e^{\mathcal{E}_{\nu,m}}\,, & \text{$(\mu_i<0 \land m>0) \lor (\mu_i>0 \land m<0)$}\,,
    \end{cases}       
\ee
}}
with
\be\label{eq:E}
\mathcal{E}_{\nu,m}=-\left|2k_m\left[-\left(2\sigma_\nu^{\rm att}\frac{N_{\rm tot}}{2}+%
      \sum_{kk^{\prime}}\int_0^{N_{\rm tot}/2} \sigma_{\nu,kk^\prime} x_k(N')\ud N'\right)\right]\right|
\ee
and
\be
  Q_i =
    \begin{cases}
      I_\nu^-(\mu_i)-\sum_{\ell=0}^L (2\ell+1)P_\ell(\mu_i)\sum_{m=1}^M R_{\ell,-m}\hat{Z}_{\nu,m}(N_{\rm tot}), & \text{$\mu_i>0$}\\
      I_\nu^+(\mu_i)-\sum_{\ell=0}^L (2\ell+1)P_\ell(\mu_i)\sum_{m=1}^M R_{\ell m}\hat{W}_{\nu,m}(0), & \text{$\mu_i<0$}\,.
    \end{cases}       
\ee
The sums in Eq.~(\ref{eq:S0}) and (\ref{eq:E}) are performed over all transitions 
contributing to the emission at frequency $\nu$.
In Eq.~(\ref{eq:E}), $I_\nu^-$ and $I_\nu^+$ represent the external field impinging on the two sides of the cloud.
In most cases, one considers a semi-infinite cloud. 
This happens, for example, in astrochemical databases where photodissociation and photoionisation rates 
are given as a function of the visual extinction or the column density. 
In this case $J_\nu(N)=f_{\nu,0}^-(N)$, while 
for clouds irradiated from two sides, $J_\nu(N)=f_{\nu,0}^-(N) + f_{\nu,0}^+(N)$.

\section{Ultraviolet photon spectrum: Results}
\label{sec:UVresults}

In this section, we show the results for typical conditions in the densest regions of molecular clouds, characterised by 
a temperature of 10~K and a turbulent line broadening of 1~km~s$^{-1}$. 
These two parameters determine the Gaussian profile, to be combined with the natural Lorentzian linewidth.
We assumed that ortho-para conversions due to reactive collisions with protons are not frequent in molecular clouds \citep{FlowerWatt1984}
and we examined the two extreme cases where the \ce{H2} o:p ratio is equal to 0:1 (\ce{H2} in para form) and to 1:0 (\ce{H2} in ortho form).
Under this hypothesis, since the ortho and para states in this framework are not coupled by any process,  
the results for the para-\ce{H2} and ortho-\ce{H2} cases (presented in Sect.~\ref{sec:photorates} and Appendix~\ref{app:photorates45}) can be linearly combined for any arbitrary o:p ratios.

\begin{figure}
\includegraphics[width=0.5\textwidth]{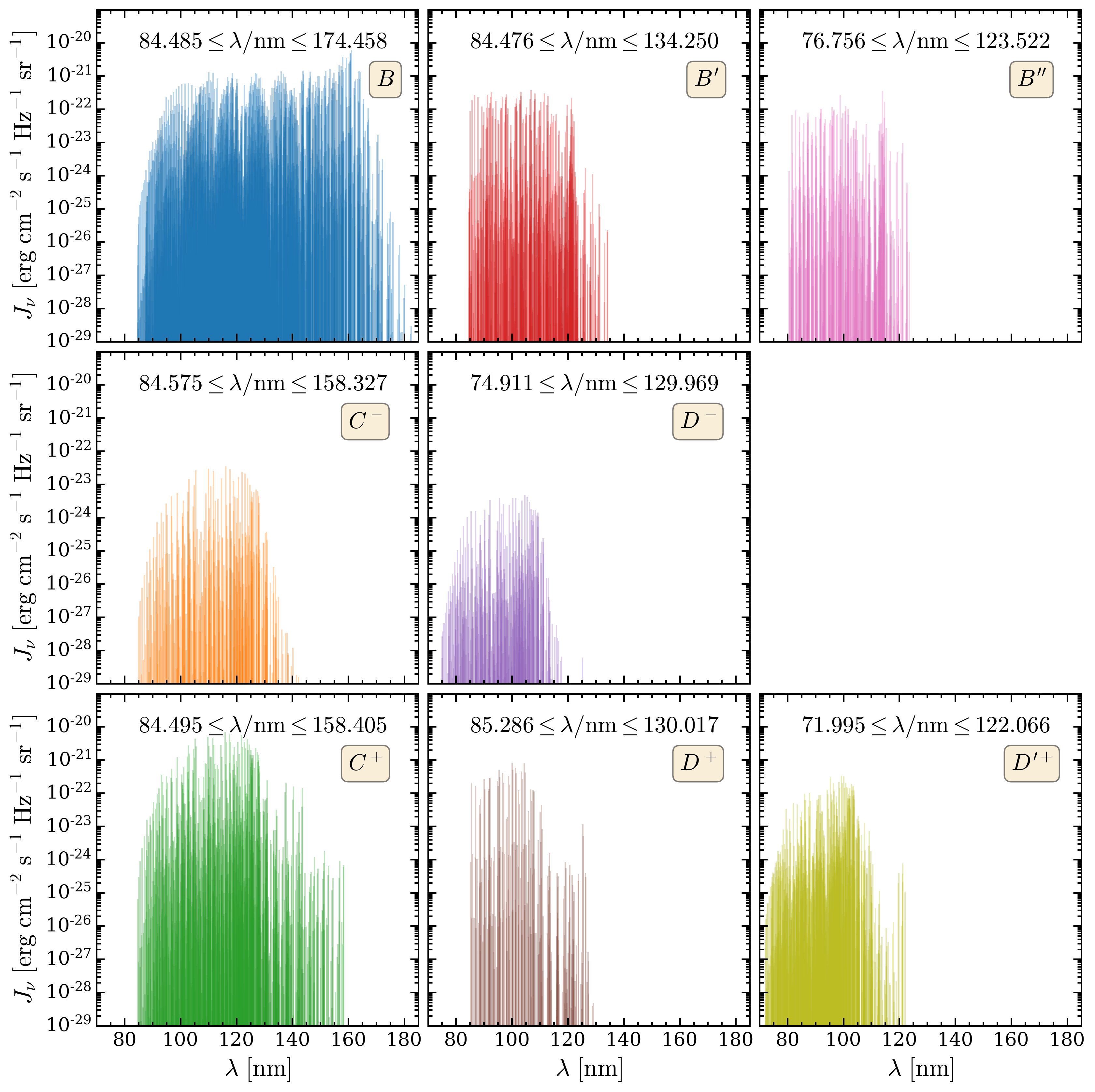}
\caption{Mean intensity, $J_\nu$, of bound-bound \ce{H2} transitions
as a function of the wavelength, $\lambda$, for a molecular cloud 
with column density $N=10^{23}$~cm$^{-2}$
illuminated by one side. 
Line spectra are shown separately for each excited electronic state.  
Results are shown for model $\mathscr{H}$, \ce{H2} o:p=0:1, and $R_V=3.1$.}
\label{fig:H2UVlinespara}
\end{figure}

\begin{figure}
\includegraphics[width=0.5\textwidth]{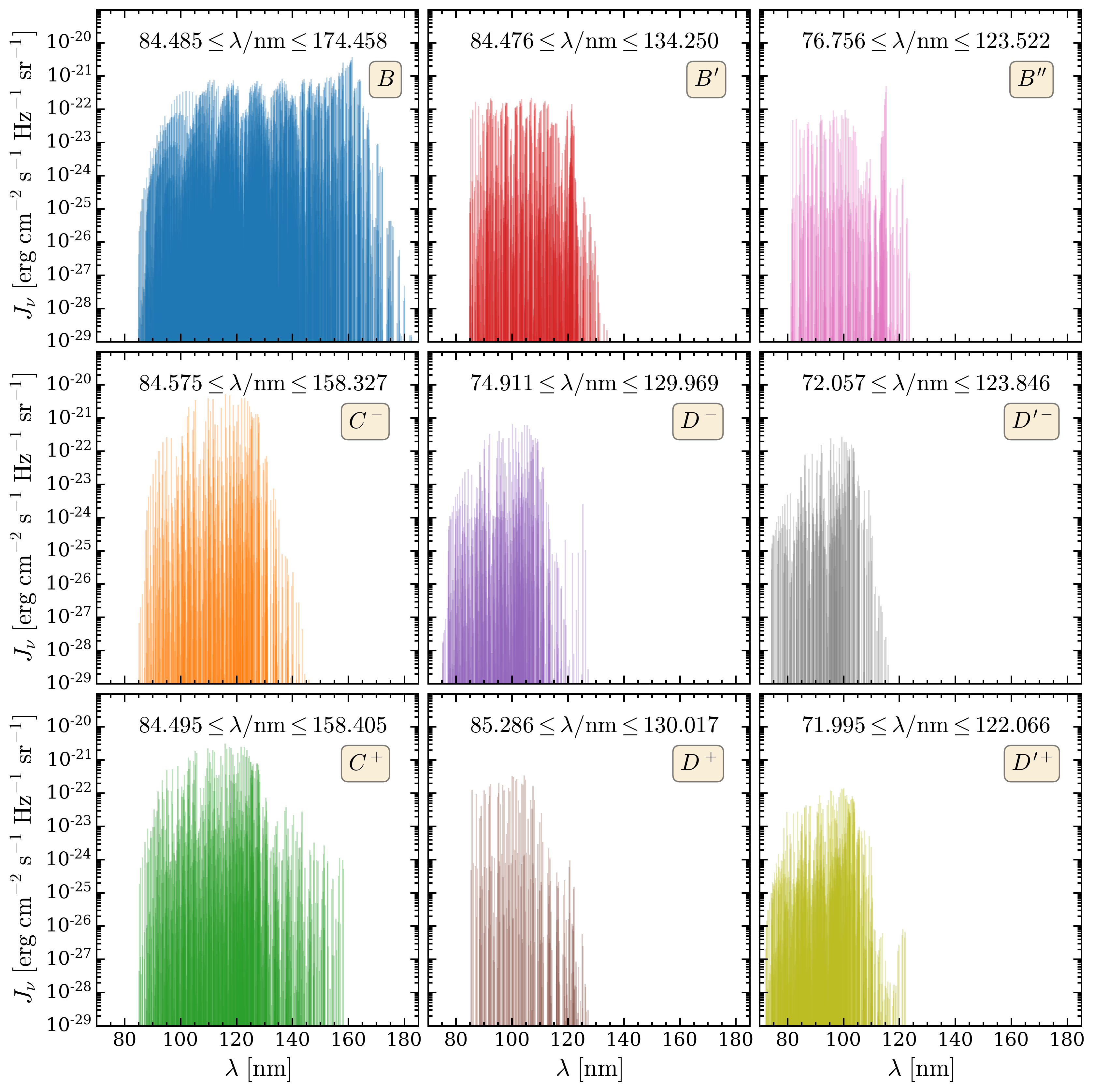}
\caption{Same as Fig.~\ref{fig:H2UVlinespara}, but for \ce{H2} o:p=1:0.}
\label{fig:H2UVlinesortho}
\end{figure}

Figures~\ref{fig:H2UVlinespara} and~\ref{fig:H2UVlinesortho} show the para-\ce{H2} and ortho-\ce{H2} line spectra, respectively. 
Each panel illustrates the partial spectrum due to the transitions from an excited electronic state ($B,C,B^\prime,D,B^{\prime\prime},D^\prime$) 
to the ground state $X$. 
As an illustration, results are shown for $N=10^{23}$~cm$^{-2}$ assuming the CR proton flux of model $\mathscr{H}$ and 
$R_V=3.1$. 

Although spontaneous emission rates for $B^{\prime\prime}\rightarrow X$ and $D^\prime\rightarrow X$ transitions are currently 
available only for a limited number of rotational levels (see Sect.~\ref{sec:energylevels}), their contribution, together with that of the $D^-\rightarrow X$ transitions, is very relevant as they generate significant emission at wavelengths shorter than 
$\sim 85$~nm,
roughly corresponding to the lower boundary of the Lyman-Werner bands, where the peak of most 
photoionisation cross sections is located \citep{Heays+2017,Hrodmarsson+2023}.
 
We note that, because of the lack of spontaneous emission rates from $J>1$ levels of the $D^{\prime-}$ state, 
in the case of pure para-\ce{H2} (Fig.~\ref{fig:H2UVlinespara}), we did not consider the $D^{\prime-}$ state, as it could not be de-excited 
if populated. Therefore, there is no associated emission.
In a future article, we will consider the emission from all upper rotational levels of the $B^{\prime\prime}$ and $D^\prime$ states.

In the case of pure ortho-\ce{H2} (Fig.~\ref{fig:H2UVlinesortho}), the emission from $D^{\prime-}$ is also present together with the emission resulting from the $Q$ branch of $C$ and $D$ states, namely $C^-$ and $D^-$. 
Moreover, the emission from the $C^-$ and $D^-$ states is much more intense than in the case of pure para-\ce{H2}. 
This is because electronic states with $\Lambda=1$, where $\Lambda$ is 
the projection of the electron
orbital angular momentum onto the internuclear axis, do not have $J=0$ levels.
Thus, if all the \ce{H2} is 
initially in the $J=0$ level (pure para-\ce{H2}), a large fraction of the emission is missing.

\begin{figure}
\includegraphics[width=0.5\textwidth]{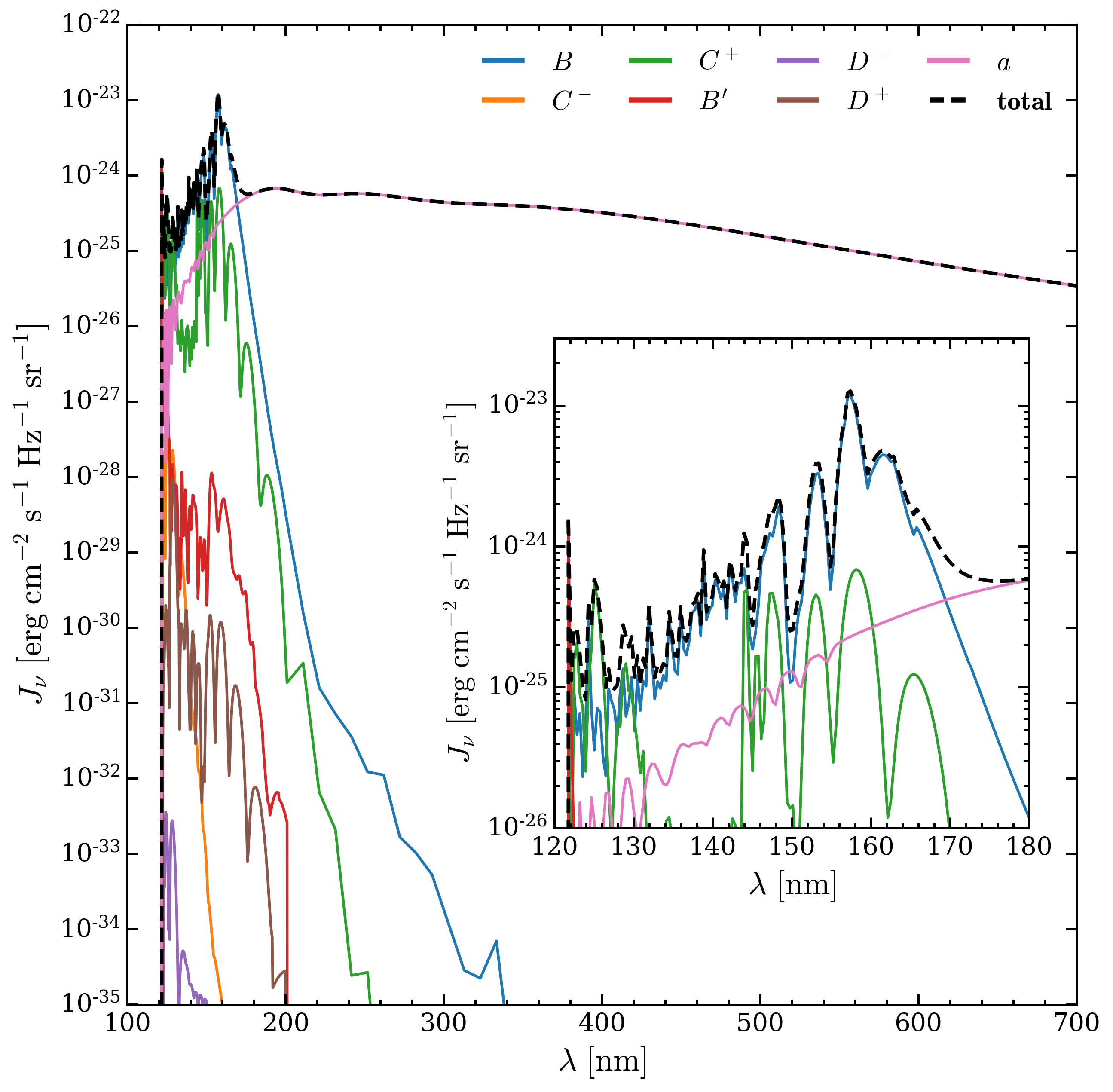}
\caption{Mean intensity, $J_\nu$, of the continuum emission of \ce{H2} from excited electronic states 
as a function of wavelength, $\lambda$, for a molecular cloud 
with a column density $N=10^{23}$~cm$^{-2}$ illuminated by
one side. The contribution of each excited electronic state (solid lines) 
is shown by solid coloured lines, while 
the dashed black line shows the total continuum emission. The inset shows the emission in the range 120--180\;nm.
Results are shown for model $\mathscr{H}$, \ce{H2} o:p=0:1, and $R_V=3.1$.}
\label{fig:H2UVcontinuumpara}
\end{figure}

\begin{figure}
\includegraphics[width=0.5\textwidth]{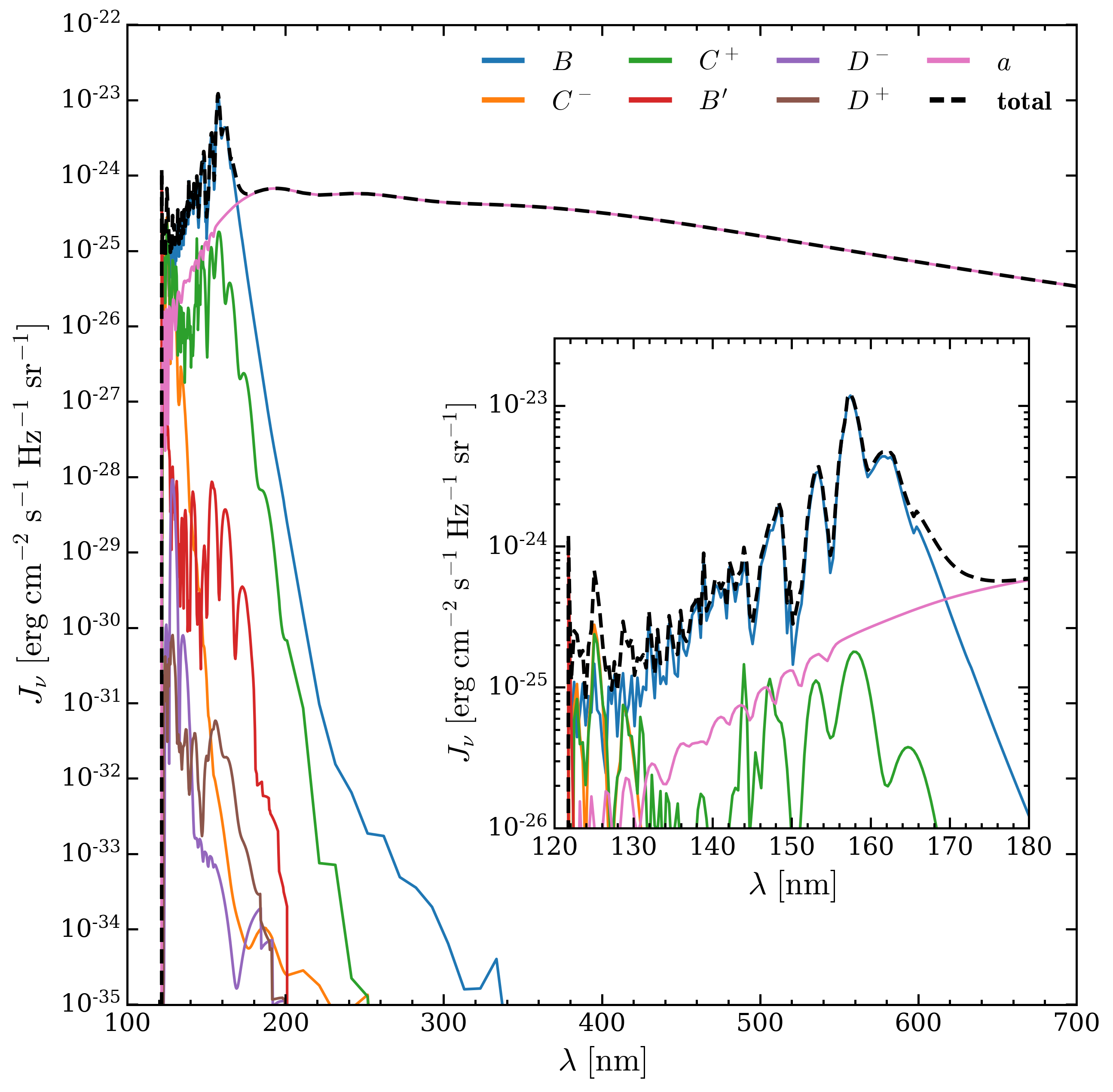}
\caption{Same as Fig.~\ref{fig:H2UVcontinuumpara}, but for \ce{H2} o:p=1:0.}
\label{fig:H2UVcontinuumortho}
\end{figure}

Figures~\ref{fig:H2UVcontinuumpara} and~\ref{fig:H2UVcontinuumortho} show the continuum spectra due to each
excited electronic state for \ce{H2} o:p=0:1 and \ce{H2} o:p=1:0, respectively.
The continuum emission is only longward of about 120~nm.
Continuum emission from $B^{\prime\prime}$ and $D^\prime$ states is not included because spontaneous 
emission rates are not available. 
We note, however, that the contributions to the continuum emission of Rydberg states with $n\geq3$ 
are more and more negligible compared to those from $n=2$ states.
Below 180~nm the continuum is dominated by the emission of the $B$ and $C^+$ states in the case o:p=0:1, 
while in the case o:p=1:0 the $C^-$ state also contributes in a small wavelength window, 
below 130 nm. 

Above 180 nm, the continuum emission is entirely due to the triplet $a$ state.
Including this emission is crucial as several photodissociation cross sections have their maximum contribution 
at $\lambda>180$~nm.
To give a few examples: \ce{NH3} shows a large number of resonances up to about 220\;nm; 
\ce{C3H3}, \ce{AlH}, and \ce{CS2} peak around 200\;nm; \ce{LiH} around 270\;nm; the \ce{S2} threshold is at about 240\;nm; \ce{CH3NH2} has a tail up to 250\;nm; and \ce{C2H5} has an important contribution above 200\;nm
\citep{Heays+2017,Hrodmarsson+2023}.

\begin{figure}
\includegraphics[width=0.5\textwidth]{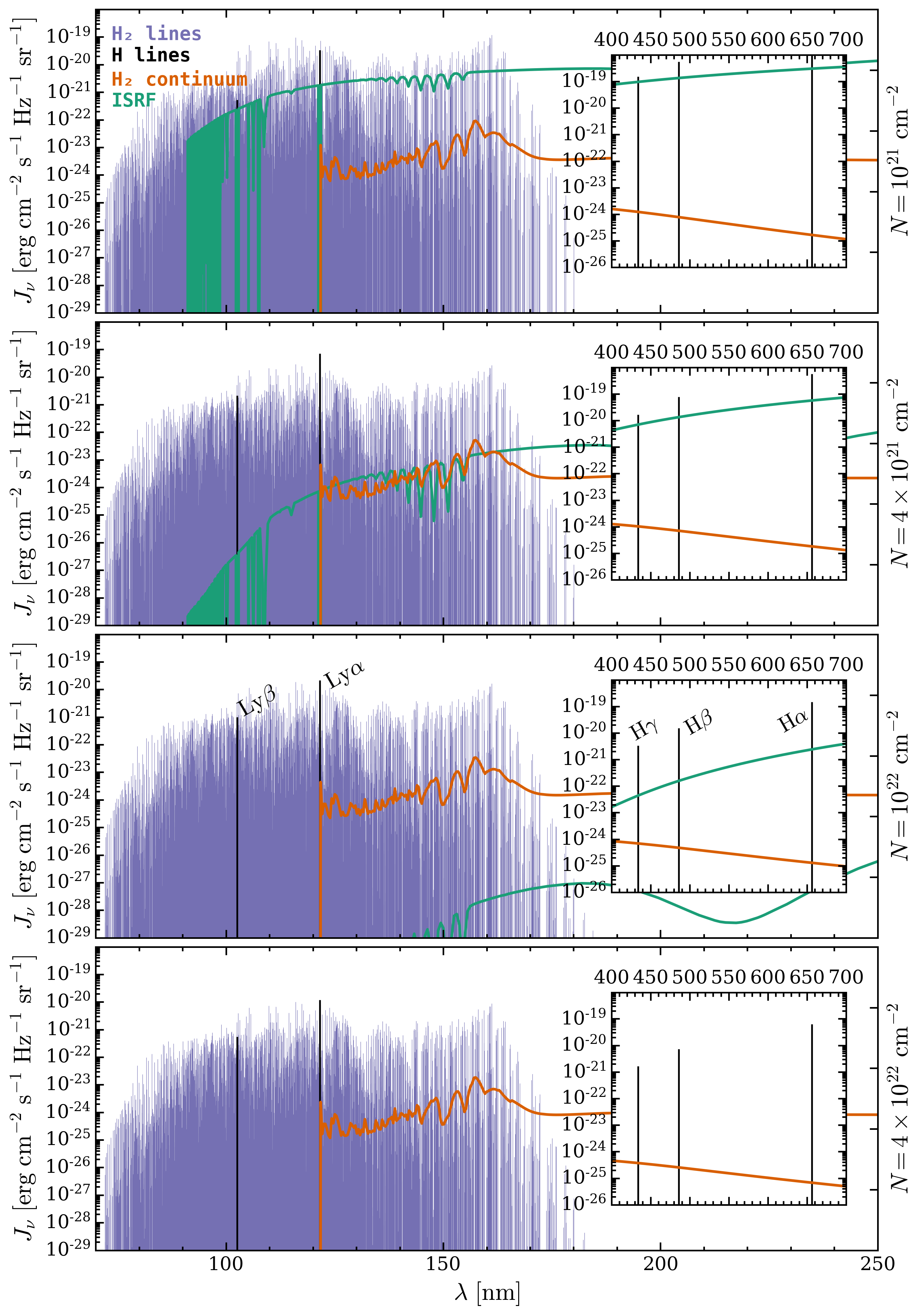}
\caption{Mean intensity, $J_\nu$, of the total emission  
in the wavelength range $72<\lambda/{\rm nm}<250$ for a molecular cloud illuminated by
one side. From top to bottom the four panels show the expected mean intensity from \ce{H2} lines, $\ce{H2}$
continuum, \ce{H} lines (Ly$\alpha$ and Ly$\beta$), and ISRF at four different column densities, from $N=10^{21}$ to $4\times10^{22}$~cm$^{-2}$.
Insets show the components of $J_\nu$ at $400<\lambda/{\rm nm}<700$, highlighting three lines of the Balmer series (H$\alpha$, H$\beta$, and H$\gamma$).
Results are shown for model $\mathscr{H}$, $\chi=1$, \ce{H2} o:p=0:1, and $R_V=3.1$.}
\label{fig:spectrumexample}
\end{figure}

Finally, in Fig.~\ref{fig:spectrumexample} we show the total specific 
intensity $J_\nu$, including \ce{H2} line and continuum components,
H line emission in the Lyman and Balmer series, and the ISRF continuum.
We show spectra at four different column densities. The comparison illustrates that the ISRF continuum becomes 
negligible compared to the line emission around $4\times10^{21}$\;cm$^{-2}$. 
However, the ISRF still dominates at wavelengths above about 250\;nm up to \ce{H2} column densities of the order of 
$10^{22}$\;cm$^{-2}$, when the \ce{H2} continuum becomes prominent.  
Atomic hydrogen lines of the Balmer series contribute significantly to the total photon flux 
(see Sect.~\ref{sec:UVflux}). 

\section{Photorates}
\label{sec:photorates}

In this section we present the calculation of photodissociation and photoionisation rates. 
This topic has already been extensively examined in previous articles thanks to the crucial advances in the 
derivation of the photodissociation and photoionisation cross sections of a large number of
atomic and molecular species, 
as presented in the Leiden Observatory database$^{\ref{fnote:photoratedatabase}}$, 
and described by \citet{Heays+2017} and by a recent update by \citet{Hrodmarsson+2023}. 
Previous calculations adopted the \ce{H2} emission
spectrum calculated by \citet{Gredel+1987,Gredel+1989}. 
As a result of recent studies on primary CR propagation \citep[e.g.][]{Padovani+2009,Padovani+2018a,Padovani+2022}, 
secondary electron generation \citep{Ivlev+2021}, rovibrationally resolved collisional excitation cross sections 
\citep{Scarlett+2023}, and spontaneous emission
rates \citep{Abgrall1993a,Abgrall1993b,Abgrall1993c,Abgrall1997,Abgrall2000,Liu+2010,Roueff2019},
we have been able to calculate the UV photon spectrum in its line and continuum components 
(Sects.~\ref{sec:UVmethod} and~\ref{sec:UVresults}), and consequently the photorates, with higher accuracy.

The rate of photodissociation or photoionisation, due to CR-generated UV photons, of an atomic or molecular species $s$ as a function of \ce{H2} column density is
\be
k_s^{\rm ph}(N)=4\pi\int \frac{J_\nu(N)\sigma_s^{\rm ph}(\nu)}{h\nu}\ud\nu\,,
\label{eq:photorates}
\ee
where $J_\nu(N)$ is the mean intensity (Eq.~\ref{eq:Jnu}) and $\sigma_s^{\rm ph}(\nu)$ is the photodissociation or 
photoionisation cross section of the species. 
We note that $J_\nu(N)$ takes into account both Doppler broadening and the natural Lorentzian linewidth, 
so that a Voigt profile is associated with each line. The number of steps for the 
frequency integration is chosen to have a resolution of at least one-hundredth of the maximum resolution of the cross sections 
for a given frequency. 
This adaptive grid has been adopted to account for all details of cross sections.

The dependence on column density lies solely in the assumption of the incident CR spectrum through the 
CR ionisation rate, $\zeta_{\ce{H2}}(N)$, parameterised as in Eq.~(\ref{eq:fittingformula}). 
Therefore, Eq.~(\ref{eq:photorates}) can be rewritten as
\be\label{eq:k0}
k_s^{\rm ph}(N) = k^{\rm ph}_{0,s} \zeta_{\ce{H2}}(N)\,.
\ee
Equation~(\ref{eq:k0}) does not include the contribution to the photodissociation or photoionisation rate of UV photons from the ISRF, 
which must be added separately.
The ISRF contribution is usually parameterised as a function of the visual extinction, whose scaling with \ce{H2} column density is given
by the standard proportionality \citep[e.g.][]{Bohlin+1978}.
Values of $k^{\rm ph}_{0,s}$ have been computed for a number of species
for $R_V=3.1$ (Tables~\ref{tab:pd3.1} and~\ref{tab:pi3.1}),
$R_{V}=4.0$ (Tables~\ref{tab:pd4.0} and~\ref{tab:pi4.0}), 
and $R_{V}=5.5$ (Tables~\ref{tab:pd5.5} and~\ref{tab:pi5.5}).
While photorates show no appreciable changes from $R_V=3.1$ to $R_V=4.0$, for $R_V=5.5$ they increase on average by $29\%\pm14\%$ and $41\%\pm9\%$ for
photodissociation and photoionisation, respectively, with respect to $R_V=3.1$.
We use the superscripts `pd' and `pi' 
to refer to the photodissociation and photoionisation constant ($k_{0,s}^{\rm pd}$ and $k_{0,s}^{\rm pi}$, respectively) 
in Eq.~(\ref{eq:k0}).
For each of these tables, we tabulate the rates for pure para and pure ortho \ce{H2} form
(o:p=0:1 and o:p=1:0, respectively). As anticipated in Sect.~\ref{sec:UVresults}, since we assume no ortho-para
conversion \citep{FlowerWatt1984}, photorates for para and ortho \ce{H2} can be linearly combined, if the o:p ratio as a function of \ce{H2} column density is known.
However, the values of $k^{\rm ph}_0$ are comparable on average to within 5\% for the two \ce{H2} o:p ratios considered, with only a few species showing slightly larger differences. 
In particular, $k_{0,s}^{\rm pd}$ for o:p=0:1 is larger than that for o:p=1:0 
by 85\% for \ce{CO}, by 30\% for \ce{H3+}, and by 25\% for \ce{C2H2} and \ce{CS}
and is lower than that for o:p=1:0 by 25\% for \ce{HCO+} and \ce{N2};
$k_{0,s}^{\rm pi}$ for o:p=0:1 is larger than that for o:p=1:0
by 40\% for \ce{CO} and \ce{Zn}, by 35\% for \ce{N2}, by 30\% for \ce{CN}, 
and by 25\% for \ce{O2}.

We compared the new photorate constants, $k^{\rm ph}_{0,s}$, 
with the most recent determination by \citet{Heays+2017} and \citet{Hrodmarsson+2023},
who computed the rates assuming $\zeta_{\ce{H2}}=10^{-16}$\;s$^{-1}$, $R_V=3.1$, o:p=0:1, a Doppler broadening of 1\;km~s$^{-1}$, and
constant abundances for H, CO, and \ce{N2} ($x_{\rm H}=10^{-4}$, $x_{\rm CO}=x_{\rm N_2}=10^{-5}$). 
Our photodissociation and photoionisation rates are, on average, smaller than the previous ones by a factor of $2.2\pm0.8$ and $1.6\pm0.5$, respectively.
For species such as \ce{AlH,C2H2,C2H3,C3H3,LiH,N2,NaCl,NaH,O_2^+,S2,SiH,l-C4}, and \ce{l-C5H}, our photodissociation rates are lower by a factor of 3 up to 5.
The photoionisation rates of \ce{H2, HF}, and \ce{N2} deserve special attention. 
For these species, the ionisation threshold energies are 15.43~eV \citep{Shiner+1993}, 16.06~eV \citep{Bieri+1980}, and 15.58~eV \citep{Trickl+1989},  
corresponding to wavelengths of about 80.36~nm, 77.19~nm, and 79.56~nm, respectively. 
This means that the contribution of photons at shorter wavelengths is crucial for the correct evaluation of the photoionisation rate. 
Our new calculations extend the photon spectrum down to 72~nm. Within this wavelength range lies a considerable number of transitions from the electronic levels $D^-$, $B^{\prime\prime}$, $D^{\prime-}$, and $D^{\prime+}$ to the ground state $X$, 
resulting in an increase in the photoionisation constants, $k_{0,s}^{\rm pi}$ of approximately $9.4\times10^3$, $2\times10^5$, and $1.2\times10^4$ for \ce{H2, HF}, and \ce{N2}, respectively.
For the \ce{H2} photodissociation rate, we also find a value larger by a factor of 16 than what was previously found. 
Again, this is because the \ce{H2} photodissociation cross section extends down to wavelengths of 70~nm and partly because of our line-by-line treatment
of \ce{H2} self-absorption.

\section{Integrated ultraviolet photon flux}
\label{sec:UVflux}
The CR-generated UV photon flux has drastic consequences on the process of charging of dust grains \citep[see e.g.][]{Ivlev+2015,Ibanez+2019}.
Through the photoelectric effect, photons can extract electrons from dust grains, thus redistributing the charge. 
More specifically, this process generates a population of positively charged grains that can then combine with those of opposite charge to form larger and larger conglomerates. 
This mechanism is particularly relevant as it is linked to the formation of planetesimals and thus planets.

\begin{figure}
\includegraphics[width=0.5\textwidth]{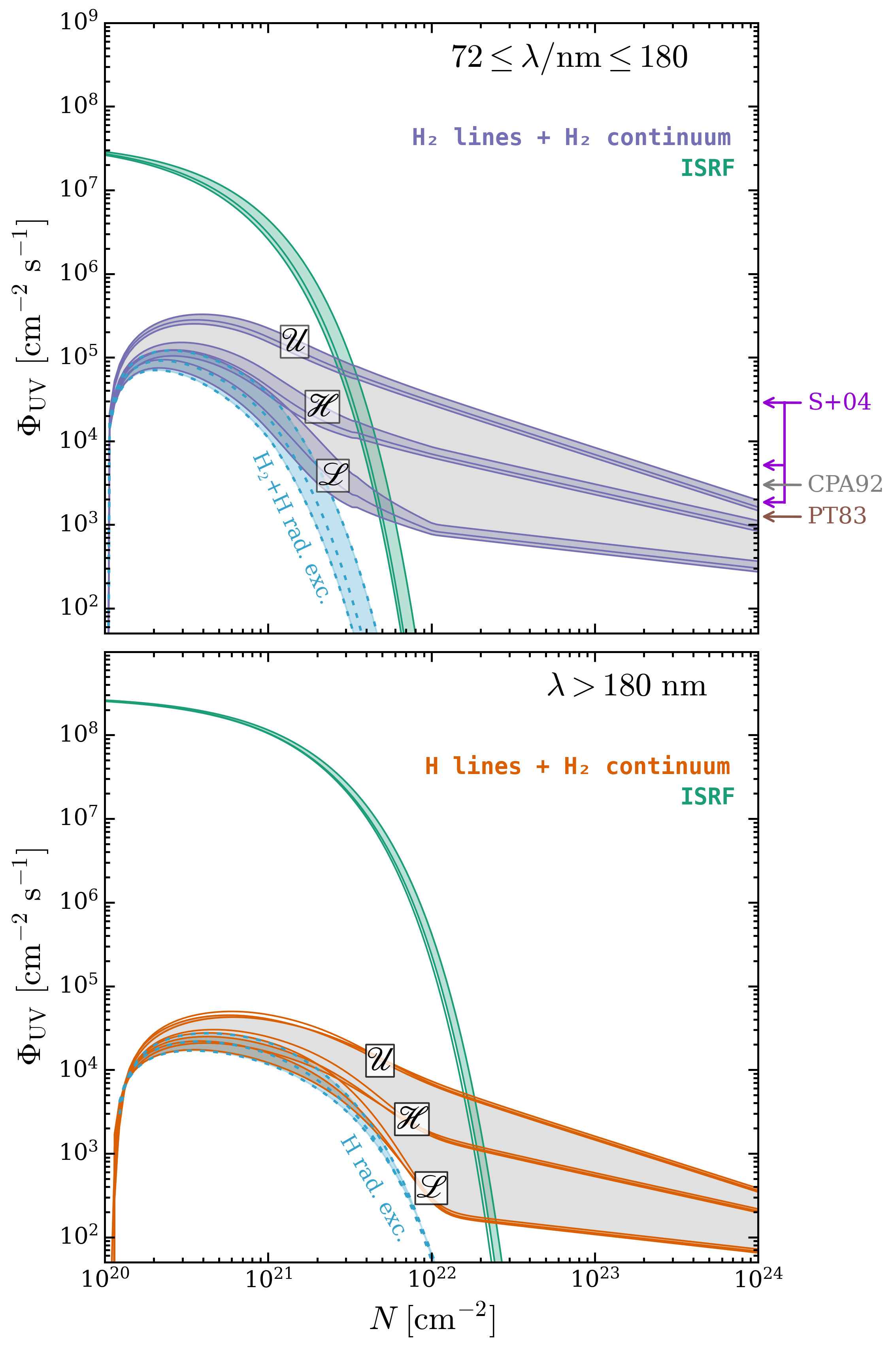}
\caption{Photon fluxes, $\Phi_{\rm UV}$, as a function of the \ce{H2} column density, $N$, 
below and above 180~nm (upper and lower panel, respectively)
for $\chi=1$.
Both panels show the integrated flux of CR-generated UV photons as a function of the three models of CR proton flux
($\mathscr{L},\mathscr{H}$ and $\mathscr{U}$) and the ISRF UV flux.
The cyan dotted lines show the integrated photon flux from \ce{H2} and H lines generated by collisional radiative excitation (labelled `rad. exc.').
The three curves for each model correspond to $R_V=3.1$ (lower curve), 4.0 (intermediate curve), and (5.5 upper curve).
Arrows on the righthand side of the upper panel show previous estimates of $\Phi_{\rm UV}$, independent of column density:
\citet[][PT83]{PrasadTarafdar1983}, \citet[][CPA92]{Cecchi-PestelliniAiello1992},
and \citet[][S+04]{Shen+2004}.}
\label{fig:Nph}
\end{figure}

The fundamental parameter governing the equilibrium charge distribution is the integrated photon flux, $\Phi_{\rm UV}$ (Eq.~\ref{eq:Nph}).
In Fig.~\ref{fig:Nph}, we show the calculation of $\Phi_{\rm UV}$ in two wavelength ranges, below and above 180~nm. 
This value is chosen to isolate the contribution of the \ce{H2} lines ($72\lesssim\lambda/{\rm nm}\lesssim180$; see Figs.~\ref{fig:H2UVlinespara} and~\ref{fig:H2UVlinesortho}). 
As can be seen, $\Phi_{\rm UV}$ depends on the selected CR model, while it has a weak dependence on the chosen $R_V$ value. 
In addition, $\Phi_{\rm UV}$ is independent of the \ce{H2} o:p ratio since it is an integrated quantity, whereas the effect of the \ce{H2} o:p ratio 
is only to redistribute photons at different wavelengths.
The bump seen between $N=10^{20}$ and $10^{22}$\;cm$^{-2}$ is due to radiative excitation alone, while at higher column density CR-generated UV photons
fully determine $\Phi_{\rm UV}$.
Photons at wavelengths larger than 180\;nm contribute as much as 20\% to the total integrated photon flux above $10^{22}$\;cm$^{-2}$.
Besides, the Lyman and Balmer series lines contribute only 3\% at $\lambda<180$~nm and up to 40\% at $\lambda>180$~nm, respectively, to $\Phi_{\rm UV}$.

As for the photorates, it turns out that $\Phi_{\rm UV}$ normalised to $\zeta_{\ce{H2}}$ 
is also independent of the assumed interstellar CR model. 
This allows us to parameterise the integrated photon flux solely as a function of column density and $R_V$ as
\be
\Phi_{\rm UV}(N) = 10^3(c_0+c_1R_V+c_2R_V^2)\left[\frac{\zeta_{\ce{H2}}(N)}{10^{-16}~{\rm s}^{-1}}\right]~{\rm cm^{-2}~s^{-1}}\,,
\label{eq:Nphfit}
\ee
where $c_0=5.023$, $c_1=-0.504$, and $c_2=0.115$. The above fit is valid for $3.1\leq R_V\leq5.5$. 
Parameterisations of $\zeta_{\ce{H2}}(N)$ are given by Eq.~(\ref{eq:fittingformula}).
In principle, Eq.~(\ref{eq:Nphfit}) is valid above about $N=10^{22}$~cm$^{-2}$; however, it can be extrapolated at lower \ce{H2} column densities as well,
as the photon flux generated by CRs is negligible with respect to the UV ISRF photons.

The upper panel of Fig.~\ref{fig:Nph} also shows 
the predictions by \citet{PrasadTarafdar1983} and \citet{Cecchi-PestelliniAiello1992} who
obtained $\Phi_{\rm UV}=1350$~cm$^{2}$~s$^{-1}$ and $\approx3000$~cm$^{2}$~s$^{-1}$, respectively,
assuming $\zeta_{\ce{H2}}=1.7\times10^{-17}$~s$^{-1}$, and by \citet{Shen+2004} who
predict constant $\Phi_{\rm UV}$ values between $1800$ and $29000$~cm$^{2}$~s$^{-1}$,
depending on the assumption of three different low-energy CR fluxes, but without accounting for energy losses while CRs propagate.

\section{Conclusions}
\label{sec:conclusions}
Cosmic rays exert significant influence by regulating the abundance of ions and radicals, 
leading to the formation of increasingly complex molecular species and influencing the charge distribution on dust grains.
Our study emphasises the critical importance of investigating the UV photon flux generated by CR secondary electrons in dense molecular clouds.
Building on the seminal works of \citet{Roberge1983}, \citet{PrasadTarafdar1983}, \citet{Sternberg1987}, and \citet{Gredel+1987,Gredel+1989}, 
we examined this topic in the light of significant advances in the field of microphysical processes. 
Our research benefited from the following advances:
$(i)$ accurate calculations of collisional excitation cross sections \citep{Scarlett+2023} and
spontaneous emission rates \citep[][Glass-Maujean, priv. comm.]{Abgrall1993a,Abgrall1993b,Abgrall1993c,Abgrall1997,Abgrall2000,Liu+2010,Roueff2019}, all of which are rotationally resolved;
$(ii)$ comprehensive insights into the propagation and attenuation of the Galactic CR flux within molecular clouds \citep{Padovani+2009,Padovani+2018a,Padovani+2022};
and $(iii)$ the robust calculation of secondary electron fluxes resulting from the ionisation of \ce{H2} by CRs \citep{Ivlev+2021}.

We were then able to calculate the population of the $X$ ground state, the excited electronic singlet ($B,C,B^\prime,D,B^{\prime\prime}$, and $D^\prime$) and triplet ($a$) states of 
molecular hydrogen, and thus the UV spectrum resulting from \ce{H2} de-excitation. 
We note that this spectrum also includes Lyman and Balmer series lines of atomic hydrogen.
From the 1508 rovibrational levels, we produced a UV spectrum consisting of 38,970 lines, spanning from 72 to 700~nm, and studied its variation as a function of 
the CR spectrum incident on a molecular cloud, 
the column density of \ce{H2}, 
the isomeric \ce{H2} composition, 
and the properties of the dust. 

Using the most recent and complete databases of photodissociation and photoionisation cross sections of a large number of atomic and molecular species
\citep{Heays+2017,Hrodmarsson+2023}, we calculated the photodissociation and photoionisation rates, giving parameterisations as a function of the \ce{H2}
column density through the CR ionisation rate models from \citet{Padovani+2022}.
On average, this new set of rates differs from previous estimates, with reductions of approximately $2.2\pm0.8$ and $1.6\pm0.5$ for photodissociation and photoionisation, respectively. 
In particular, deviations of up to a factor of 5 are observed for some species, such as \ce{AlH, C2H2, C2H3, C3H3, LiH, N2, NaCl, NaH, O2+, S2, SiH, l-C4}, and \ce{l-C5H}.
Particular consideration should be paid to the significantly higher photoionisation rates of \ce{H2}, \ce{HF,} and \ce{N2}, 
as well as the photodissociation of \ce{H2}, which our study revealed to be orders of magnitude higher than previous evaluations. 
This discrepancy can be attributed to our new calculations extending the photon spectrum down to 72~nm, where the cross sections of these species have a large contribution,
and partly to the \ce{H2} self-absorption that we were able to treat line by line.

In addition, we calculated the integrated UV photon flux, 
regulating the equilibrium charge distribution on dust grains 
and thus the formation of larger and larger conglomerates.
Compared to previous estimates of this quantity that predicted a constant value,
we provide a parameterisation as a function of \ce{H2} column density, 
through the CR ionisation rate, and the dust properties.

\begin{acknowledgements}
The authors wish to thank the referee, John Black, for his careful reading of the manuscript and insightful comments that considerably helped to improve the paper. 
The authors are also grateful to Herv\'e Abgrall, Martin \v Ci\v zek, Mich\`ele Glass-Maujean, Isik Kanik, Xianming Liu, Evelyne Roueff, and Jonathan Tennyson for fruitful discussions and feedback.
I, M.P., want to leave a memory of my father Piero, a watchmaker. I remember when I was trying to explain the underlying idea of this work to you during one of our last walks, while you were telling me about how you had repaired the clock tower in Buggiano, our village.
\end{acknowledgements}

\bibliographystyle{aa} % style aa.bst
\bibliography{mybibliography-bibdesk.bib} 

\begin{appendix}

\section{Comparison with \citet{Gredel+1989}}
\label{app:comparisongredel}

\citet{Gredel+1989}, hereafter G89, assume that H$_2$ is in some $v=0$, $J=J_0$ level(s) of the ground state $X$,  
and is collisionally excited by a 30~eV secondary CR electron to a level $v^\prime J^\prime$ of an excited electronic state $S$,
from which it spontaneously decay to a level $v^{\prime\prime} J^{\prime\prime}$
of the ground state with probability $A_{S v^\prime\! J^\prime\rightarrow X v^{\prime\prime}\! J^{\prime\prime}}$.
The probability of emission of a line photon with energy $h\nu=E(S v^\prime\! J^\prime)-E(X v^{\prime\prime}\! J^{\prime\prime})$ is
\be
P_{\nu,S}=\sum_{J_0} x_{J_0} \times \frac{C^{\rm CR}_{J_0\rightarrow Sv^\prime J^\prime}}{C^{\rm CR}_{J_0\rightarrow S}}
\times \frac{A_{Sv^\prime J^\prime\rightarrow Xv^{\prime\prime}J^{\prime\prime}}}{A_{S v^\prime\! J^\prime\rightarrow X}}\,,
\ee
where $x_{J_0}$ is the fractional population of each level $v=0$, $J=J_0$ of the ground state, 
$C^{\rm CR}_{J_0\rightarrow S}$ is the total collisional excitation rate to all levels $v^\prime\! J^\prime$ 
of state $S$\footnote{G89 factorise the rotational dependence of collisional excitation rates with H\"onl-London factors.},
\be
C^{\rm CR}_{J_0\rightarrow S}=\sum_{v^\prime\! J^\prime} C^{\rm CR}_{J_0\rightarrow S v^\prime\! J^\prime}\,,
\ee
and 
$A_{S v^\prime\! J^\prime}$
is the total decay rate of the level $v^\prime\! J^\prime$ of state $S$,
\be
A_{S v^\prime\! J^\prime \rightarrow X}=\sum_{v^{\prime\prime}\! J^{\prime\prime}} A_{S v^\prime\! J^\prime\rightarrow X v^{\prime\prime}\! J^{\prime\prime}}.
\ee
The emission rate $R_\nu$ of photons of frequency $\nu$ is then the sum over all $S$ states of the emission probabilities $P_{\nu,S}$
multiplied by the total excitation rate $C^{\rm CR}_S$ of each state:
\be
R_\nu=\sum_S P_{\nu,S} C^{\rm CR}_S,
\ee
where
\be
C^{\rm CR}_S=\sum_{J_0} x_{J_0} C^{\rm CR}_{J_0\rightarrow S}.
\ee
For a statistical equilibrium mixture of H$_2$ with $x_{J_0=0}=0.25$ and $x_{J_0=1}=0.75$, G89 obtain the weights listed in Table~\ref{tabG89}. The weights are normalised to $\zeta_{{\rm H}_2}$, the total H$_2$ ionisation rate (primary plus mono-energetic secondary electrons with energy $30$~eV). Our results are shown for comparison\footnote{For excitations within state $X$, G89 find $C_X^{\rm CR}/\zeta_{{\rm H}_2}= 8.90$. We find 12.041.}.

\begin{table}[!h]
\label{tabG89}
\caption{CR excitation rates, $C^{\rm CR}_S$, normalised to the CR ionisation rate, $\zeta_{\ce{H2}}$.}
\begin{center}
\begin{tabular}{lll}
\toprule\toprule
state $S$ & \multicolumn{2}{c}{$C^{\rm CR}_S/\zeta_{{\rm H}_2}$}\\
\cmidrule{2-3}
& G89 & this work\\
\midrule
$B$ & 0.19 & 0.151\\
$C$ & 0.077 & 0.131\\
$B^\prime$ & 0.01 & 0.014\\
$D$ & 0.01 & 0.017\\
$B^{\prime\prime}$ & 0.005 & 0.004\\
$D^{\prime}$ & 0.005 & 0.011\\
$a$ & 0.03 & 0.029\\
$b$ & 0.42 & 0.378\\
\bottomrule
\end{tabular}
\tablefoot{The normalised CR excitation rates are computed separately
for each excited electronic state, $S$.}
\end{center}
\end{table}

The mean intensity of H$_2$ line emission in the optically thick limit is then
\be
J_\nu \approx \frac{h\nu}{4\pi (1-\omega)\sigma_\nu^{\rm ext}}R_\nu,
\ee
where $\sigma_\nu^{\rm ext}$ is the dust extinction cross section per H nucleus, $\omega$ the dust grain albedo, 
and absorption by gas species has been neglected\footnote{G89 assume constant values of $\sigma_\nu^{\rm ext}=2\times 10^{-21}$\;cm$^2$ and $\omega=0.5$.}. The photodissociation or photoionisation rate of a species $s$ is
\be
k_s^{\rm ph}=4\pi\int \frac{J_\nu\sigma_s^{\rm ph}(\nu)}{h\nu}\,{\rm d}\nu \approx \int \frac{R_\nu \sigma_s^{\rm ph}(\nu)}{(1-\omega)\sigma_\nu^{\rm ext}}\,{\rm d}\nu
=\frac{\zeta_{{\rm H}_2} p_s}{1-\omega}\,,
\ee
where $p_s$ is an efficiency factor defined by
\be
p_s=\frac{1}{\zeta_{{\rm H}_2}}\int \frac{R_\nu \sigma_s^{\rm ph}(\nu)}{\sigma_\nu^{\rm ext}}\,{\rm d}\nu.
\ee
Therefore, the values of $p_s$ tabulated by G89 can be compared to our values of $(1-\omega)k_{0,s}^{\rm ph}$.

\section{Update of cosmic-ray ionisation data}
\label{app:CRionobstheory}

\begin{figure}[!h]
\begin{center}
\resizebox{1\hsize}{!}{\includegraphics[]{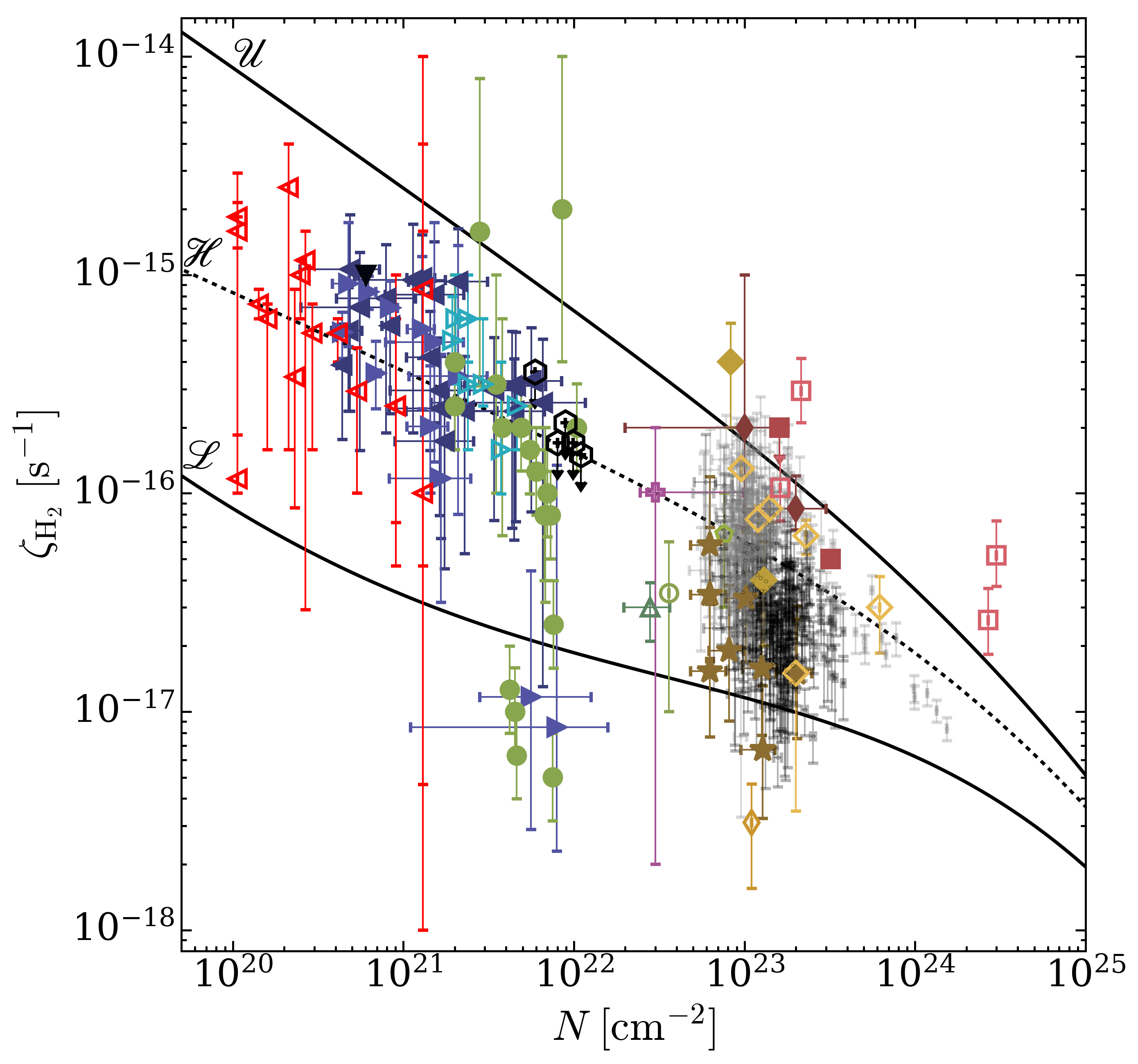}}
\caption{Total CR ionisation rate as a function of the H$_{2}$ column density. 
Theoretical models $\mathscr{L}$ (lower solid black line), $\mathscr{H}$ (dotted black line),
and $\mathscr{U}$ (upper solid black line).
Expected values from models also include the ionisation due to primary CR electrons and secondary
electrons.
Observational estimates in diffuse clouds:
solid down-pointing triangle \citep{Shaw+2008},
solid up-pointing triangle \citep{Neufeld+2010},
solid left-pointing triangles \citep{IndrioloMcCall2012},
solid right-pointing triangles \citep{NeufeldWolfire2017},
empty right-pointing triangles \citep{Luo+2023a},
empty left-pointing triangles \citep{Luo+2023b};
in low-mass dense cores:
solid circles \citep{Caselli+1998},
empty hexagons \citep{Bialy+2022},
empty circle \citep{MaretBergin2007},
empty pentagon \citep{Fuente+2016},
empty up-pointing triangle \citep{Redaelli+2021};
in high-mass star-forming regions:
stars \citep{Sabatini+2020},
black and grey small symbols \citep{Sabatini+2023},
solid diamonds \citep{deBoisanger+1996},
empty diamonds \citep{VanderTak+2000},
empty thin diamonds \citep{Hezareh+2008},
solid thin diamonds \citep{MoralesOrtiz+2014};
in circumstellar discs:
empty squares \citep{Ceccarelli+2004};
in massive hot cores:
solid squares \citep{BargerGarrod2020}.
}
\label{fig:zvsN}
\end{center}
\end{figure}

In Fig.~\ref{fig:zvsN}, 
we present the most-updated compilation of CR ionisation rate estimates obtained from observations 
in diffuse clouds, low- and high-mass star-forming regions, circumstellar discs, and massive hot cores
\citep[see also][]{Padovani2023}.
In the same plot we show the trend of $\zeta_{\ce{H2}}$ predicted by CR propagation models 
\citep[e.g.][]{Padovani+2009,Padovani+2018a,Padovani+2022} described in Sect.~\ref{sec:esecspectra}.
Models also include the contribution of primary CR electrons and secondary electrons.
We note that the models presented here only account for the propagation of 
interstellar CRs, 
but in more evolved sources, such as in high-mass star-forming regions and hot cores, 
there could be a substantial contribution from locally accelerated charged particles 
\citep{Padovani+2015,Padovani+2016,GachesOffner2018,Padovani+2021a}.

Below, we also present polynomial fits of the three trends of $\zeta_{\ce{H2}}$ predicted by the models. 
These parameterisations differ from those presented in \citet{Padovani+2018a} as here we take into account the rigorous calculation of secondary electrons 
presented in \citet{Ivlev+2021} in the regime of continuous slowing-down approximation \citep{Takayanagi1973,Padovani+2009}. 
The CR ionisation rate can be parameterised with the following fitting formula:
\be\label{eq:fittingformula}
\log_{10} \left(\frac{\zeta_{\ce{H2}}}{\rm s^{-1}}\right)=\sum_{k\ge0} c_k \log_{10}^k \left(\frac{N}{\rm cm^{-2}}\right)\,.
\ee
Equation~(\ref{eq:fittingformula}) is valid for $10^{19}\le N/{\rm cm^{-2}}\le 10^{25}$. The coefficients $c_k$ are given in Table~\ref{tab:zetafit}.

\begin{table}[!h]
\caption{Coefficients $c_k$ of the polynomial fit, Eq.~(\ref{eq:fittingformula}).}
\begin{center}
\small
\begin{tabular}{cccc}
\toprule\toprule
$k$ & model $\mathscr{L}$ & model $\mathscr{H}$ & model $\mathscr{U}$\\
\midrule
0 & $-2.93027572(2)$ & $-2.47467467(2)$ & $-2.25202661(2)$ \\
1 & \phantom{$-$}$6.02083162(1)$ & \phantom{$-$}$4.60847057(1)$ & \phantom{$-$}$4.22778765(1)$\\
2 & $-4.72413075$ & $-3.37884449$ & $-3.09907676$\\
3 & \phantom{$-$}$1.60518530(-1)$ & \phantom{$-$}$1.09332535(-1)$ & \phantom{$-$}$9.97659175(-2)$\\
4 & $-2.01065346(-3)$ & $-1.32768917(-3)$ & $-1.20564293(-3)$\\
\bottomrule
\end{tabular}
\tablefoot{The coefficients are given for the three models of interstellar CR proton spectrum 
($\mathscr{L}$, $\mathscr{H}$, and $\mathscr{U}$; see Sect.~\ref{sec:CRspectrum}).}
\normalsize
\end{center}
\label{tab:zetafit}
\end{table}%

\section{Updated energy loss functions}
\label{app:lossfunctions}

The quantity that governs the decrease in energy of a particle as it moves through a medium is referred to as the energy loss function.
For electrons colliding with \ce{H2} molecules, the energy loss function is given by
\begin{align}
\label{eq:elossfunction}
L_{e}(E)&=L_e^{\rm C}(E)\frac{2m_{e}}{m_{\ce{H2}}}\sigma^{\rm m.t.}_e(E)E+\sum_{j}\sigma^{{\rm exc},j}_e(E)E_{{\rm thr},j}+\nonumber\\
&\int_{0}^{(E-I)/2}\frac{\ud\sigma^{\ce{H2}}_e(E,\varepsilon)}{\ud\varepsilon}(I+\varepsilon)\ud\varepsilon+\nonumber\\
&\int_{0}^{E}\frac{\ud\sigma^{\rm br}_e(E,E_{\gamma})}{\ud E_{\gamma}}E_{\gamma}\ud E_{\gamma}+KE^{2}\,.
\end{align}
The terms on the right-hand side are momentum transfer, rotational and vibrational excitation, electronic excitation, ionisation, bremsstrahlung, and synchrotron losses.
Here, $m_{e}$ and $m_{\ce{H2}}$ denote the mass of the electron and \ce{H2}, respectively,
$\sigma^{\rm m.t.}_e$ and $\sigma^{{\rm exc},k}_e$ are the cross sections of momentum transfer and excitation of state $k$,
$E_{{\rm thr},k}$ is the corresponding threshold energy of the excitation, 
$\ud\sigma^{\ce{H2}}_e/\ud\varepsilon$ is the differential ionisation cross section \citep{Kim+2000},
where $\varepsilon$ is the energy of the secondary electron and $I=15.43$~eV is the ionisation threshold,
$\ud\sigma^{\rm br}_e/\ud E_{\gamma}$ is the bremsstrahlung differential cross section
\citep{BlumenthalGould1970}, where $E_{\gamma}$ is the energy of the emitted photon,
and $KE^{2}$ represents synchrotron losses with $K=5\times10^{-38}$~eV~cm$^{2}$ and $E$ in eV
\citep{Schlickeiser2002book}.
We assume the relationship 
by \citet{Crutcher2012} between the magnetic field strength, $B$, and the volume density, $n$, $B = B_0(n/n_0)^\kappa$, 
with $B_0 = 10~\mu$G, $n_0 = 150$~cm$^{-3}$, and $\kappa = 0.5-0.7$. 
Choosing $\kappa = 0.5$ removes the dependence on volume density \citep[see][for details]{Padovani+2018a}.
Depending on the ionisation fraction and the temperature of the medium, Coulomb losses, $L_e^{\rm C}(E)$, must be also included \citep{Swartz+1971}.

Similarly, for protons colliding with \ce{H2}, the energy loss function is given by
\begin{align}
\label{eq:plossfunction}
L_{p}(E)&=L_p^{\rm C}(E)+\frac{2m_{p}m_{\ce{H2}}}{(m_p+m_{\ce{H2}})^2}\sigma^{\rm m.t.}_p(E)E+\sum_{j}\sigma^{{\rm exc},j}_p(E)E_{{\rm thr},j}+\nonumber\\
&(I+\langle \varepsilon\rangle)\frac{\sigma^{\rm e.c.}_p\sigma^{\rm self-ion}_{\rm H}}{\sigma^{\rm e.c.}_p+\sigma^{\rm self-ion}_{\rm H}}+\nonumber\\
&\int_{0}^{4m_eE/m_p}\frac{\ud\sigma^{\ce{H2}}_p(E,\varepsilon)}{\ud\varepsilon}(I+\varepsilon)\ud\varepsilon+%
L_p^\pi(E)\,.
\end{align}
In addition to the momentum transfer, excitation, and ionisation terms described in Eq.~(\ref{eq:elossfunction}) for the electron energy loss function, 
on the right-hand side there are losses due to electron capture (term in the second row of Eq.~\ref{eq:plossfunction}) and pion production (last term of Eq.~\ref{eq:plossfunction}). 

Energy losses by electron capture can be derived following the methodology presented in the works by \citet{Edgar1973a,Miller1973,Edgar1975}.
The following cycle must be considered:
\begin{align*}
     p + \ce{H2} &\rightarrow \ce{H}_{\rm fast} + \ce{H2+} \\
     \ce{H}_{\rm fast} + \ce{H2} &\rightarrow p + \ce{H2} + e^-
     \,.
\end{align*}
During the first process, the ionisation energy of \ce{H2} is lost, and in the second process the average energy of the ejected electron,
$\langle\varepsilon\rangle$, is lost.
Thus, the contribution of the electron capture cycle to the loss function is 
\be
L^{\rm e.c.}_p=f_p I \sigma^{\rm e.c.}_p + f_{\rm H} \langle\varepsilon\rangle \sigma_{\rm H}^{\rm self-ion}\,, 
\ee
where 
\be
f_{\rm H} = \frac{\sigma^{\rm e.c.}_p}{\sigma^{\rm e.c.}_p+\sigma^{\rm self-ion}_{\rm H}}
\ee
and $f_p = 1-f_{\rm H}$. Here $\sigma^{\rm e.c.}_p$ is the electon capture cross section \citep{Gilbody1957,Curran1959,deHeer1966,Toburen1972,Rudd1983,Gealy1987,Baer1988,Phelps1990,Errea2010}, 
$\sigma^{\rm self-ion}_{\rm H}$ is the self-ionisation of hydrogen atoms \citep{Stier1956,VanZyl1981,Phelps1990}, and
\be
\langle\varepsilon\rangle = \varepsilon_0-\frac{\varepsilon_{\rm max}}{\exp(\varepsilon_{\rm max}/\varepsilon_0)-1}\,,
\ee
where $\varepsilon_{\rm max}=4(m_e/m_p)E$ is the maximum energy of the ejected secondary electron 
corresponding to an incident proton of energy $E$,
and $\varepsilon_0$ is the saturation energy of the secondary electron, adjusted to reproduce
the experimental data, which is set to 20~eV.
Finally, the expression for Coulomb losses, $L_p^{\rm C}(E)$, is given by \citet{MannheimSchlickeiser1994}
and for pion losses, $L_p^\pi$, whose contribution dominates above 280~MeV is given by \citet{KrakauSchlickeiser2015}.

\begin{figure*}[!h]
\begin{center}
\resizebox{.93\hsize}{!}{\includegraphics[]{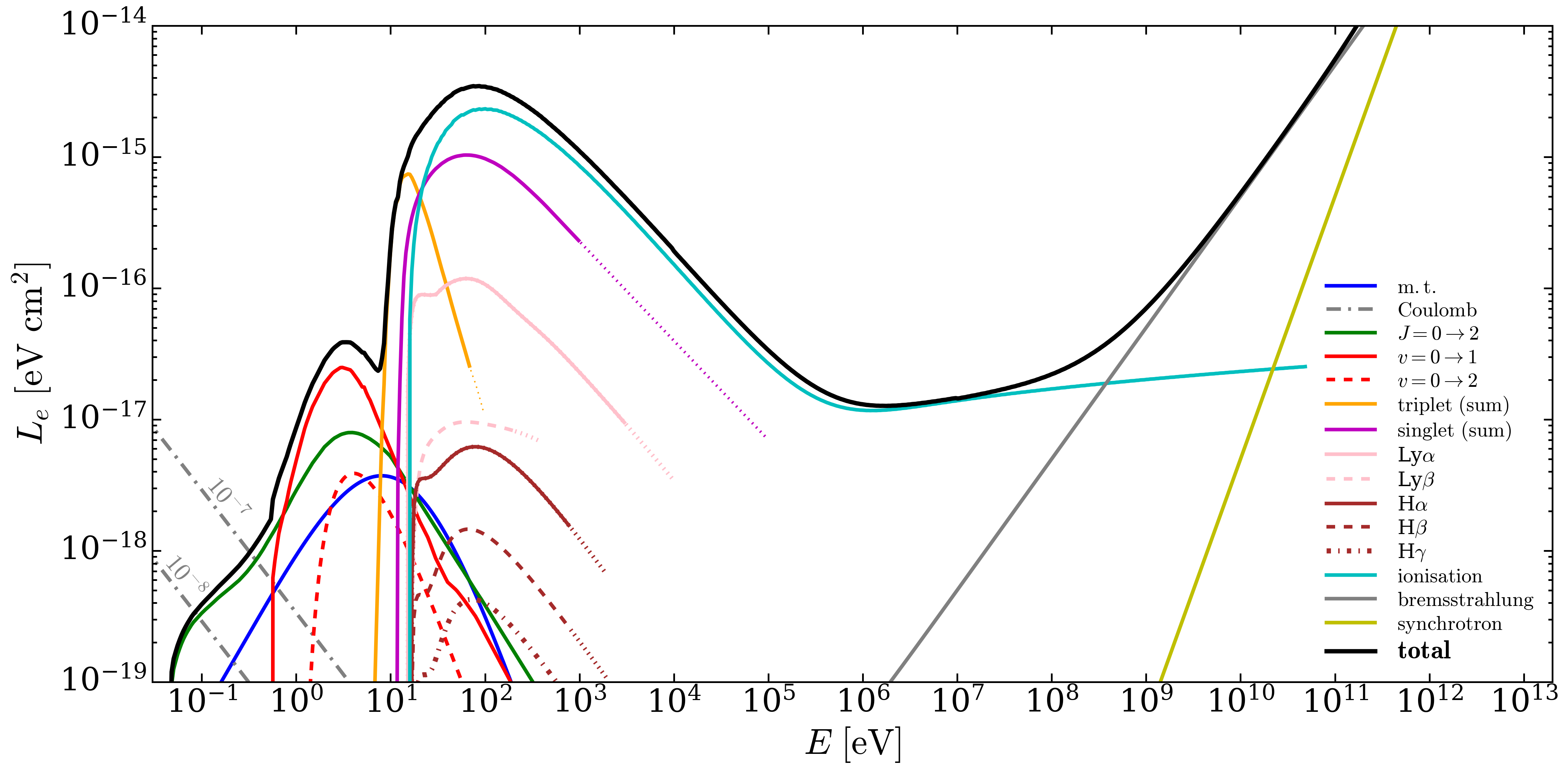}}
\caption{Energy loss function for electrons colliding with \ce{H2} (solid black line). 
Coloured lines show the different components and the following references point to the papers from which the relative cross sections have been adopted: 
momentum transfer (‘m.t.’, solid blue; \citealt{PintoGalli2008}); 
the rotational transition $J = 0 \rightarrow 2$ (solid green line; \citealt{England+1988}); 
vibrational transitions $v = 0 \rightarrow 1$ (solid red line; \citealt{Yoon+2008}) and $v = 0 \rightarrow 2$ (dashed red line; \citealt{Janev+2003}); 
electronic transitions summed over all the triplet and singlet states (solid orange and magenta lines, respectively; \citealt{Scarlett+2021a});
Lyman series (solid pink lines; \citealt{vanZyl+1989,Ajello+1991,Ajello+1996});
Balmer series (solid brown lines; \citealt{Mohlmann+1977,Karolis+1978,Williams+1982});
ionisation (solid cyan line; \citealt{Kim+2000}); 
bremsstrahlung (solid grey line; \citealt{BlumenthalGould1970}; \citealt{Padovani+2018b}); 
and synchrotron (solid yellow line; \citealt{Schlickeiser2002book}). 
Dash-dotted grey lines show the Coulomb losses at 10~K for ionisation fractions equal to $10^{-7}$ and $10^{-8}$ \citep{Swartz+1971}
and the solid yellow line shows the synchrotron losses \citep{Schlickeiser2002book}.}
\label{fig:ElosseH2}
\end{center}
\end{figure*}

\begin{figure*}[!h]
\begin{center}
\resizebox{.93\hsize}{!}{\includegraphics[]{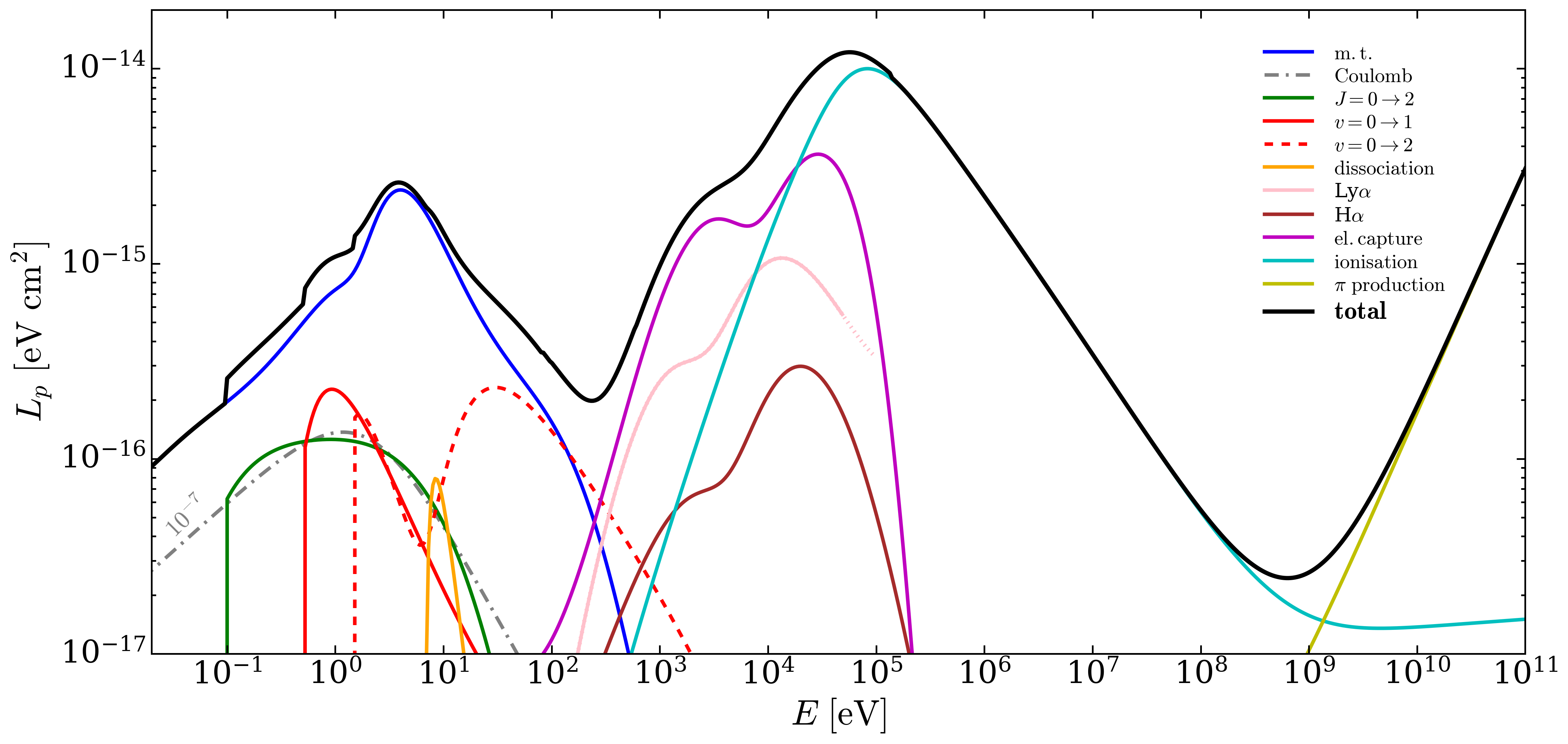}}
\caption{Energy loss function for protons colliding with \ce{H2} (solid black line). 
Coloured lines show the different components and the following references point to the papers from which the relative cross sections have been adopted: 
momentum transfer (‘m.t.’, solid blue; \citealt{PintoGalli2008}); 
the rotational transition $J = 0 \rightarrow 2$ (solid green line; \citealt{Gianturco+1977,Linder1980}); 
vibrational transitions $v = 0 \rightarrow 1$ and $v = 0 \rightarrow 2$ (solid and dashed red lines, respectively; \citealt{Gentry1975,Herrero1972,Schinke1977,Niedner1987,Janev+2003}); 
dissociation (solid orange line; \citealt{Janev+2003});
Ly$\alpha$ (solid pink line; \citealt{vanZyl1967,vanZyl+1989,Phelps1990});
H$\alpha$ (solid brown line; \citealt{Williams+1982});
electron capture (solid purple line; \citealt{Gilbody1957,Curran1959,deHeer1966,Toburen1972,Rudd1983,Gealy1987,Baer1988,Phelps1990,Errea2010});
ionisation (solid cyan line; \citealt{Rudd1988,Krause2015}); 
and pion production (solid yellow line; \citealt{KrakauSchlickeiser2015}). 
Dash-dotted grey lines show the Coulomb losses at 10~K for an ionisation fraction of $10^{-7}$ \citep{MannheimSchlickeiser1994}.}
\label{fig:ElosspH2}
\end{center}
\end{figure*}

\section{Photorates for various $R_V$ values}
\label{app:photorates45}

\begin{table*}
\caption{Photodissociation rates normalised to the CR ionisation rate, $k_{0,s}^{\rm pd}$, 
for \ce{H2} o:p=0:1 and \ce{H2} o:p=1:0 and for $R_V=3.1$.}
\begin{center}
\resizebox{\linewidth}{!}{
\begin{tabular}{lcc|lcc|lcc}
\toprule\toprule
\tcbox[sharp corners, colframe=black, colback=white, size=fbox]{$R_V=3.1$} & \multicolumn{2}{c}{$k_{0,s}^{\rm pd}$} & & \multicolumn{2}{c}{$k_{0,s}^{\rm pd}$} && \multicolumn{2}{c}{$k_{0,s}^{\rm pd}$}\\
\cmidrule{2-3} \cmidrule{5-6} \cmidrule{8-9}
Species $s$ & o:p=0:1 & o:p=1:0 & Species $s$ & o:p=0:1 & o:p=1:0 & Species $s$ & o:p=0:1 & o:p=1:0\\
\midrule
\ce{Al}         &	--		  &	--	        &	\ce{Ca+}	&	--		    &	--      	&	\ce{NH2CHO}	&	1.292(3)    & 1.294(3)      \\
\ce{AlH}        &	1.641(1)  & 1.644(1)	&	\ce{Cl}  	&	--		    &	--   	    &	\ce{NH3}	&	6.266(2)    & 6.093(2)  	\\
\ce{C}          &	--	      &	--         	&	\ce{Cr} 	&	--		    &	--  	    &	\ce{NO}	    &	1.715(2)    & 1.673(2)  	\\
\ce{C2}         &	1.368(2)  & 1.368(2)	&	\ce{Fe}    	&	--		    &	--  	    &	\ce{NO2}	&	6.022(2)    & 6.046(2)  	\\
\ce{C2H}        &	8.184(2)  & 9.006(2)	&	\ce{H}	    &	--		    &	--     	    &	\ce{Na}    	&	--	    	& --    	    \\
\ce{C2H-}       &	--		  &	--        	&	\ce{H-}    	&	--	    	&	--   	    &	\ce{NaCl}	&	5.030(1)    & 5.036(1)	    \\	
\ce{C2H2}       &	1.086(3)  & 8.684(2)	&	\ce{H2}	    &	4.048(2)    &   3.912(2)	&	\ce{NaH}	&	2.323(2)	& 2.368(2) 	    \\	
\ce{C2H3}       &	1.567(2)  & 1.566(2)    &	\ce{H2+}	&	3.238(2)	&	3.290(2)    &	\ce{Ni}    	&	--		    & --          	\\
\ce{C2H4}       &	1.450(3)  & 1.450(3)	&	\ce{H2CO}	&	6.247(2)    &   6.054(2)	&	\ce{O}   	&	--		    & --       	    \\
\ce{C2H5}       &	2.278(2)  & 2.262(2)	&	\ce{H2CS}	&	9.744(2)    &   9.734(2)	&	\ce{O2}  	&	3.077(2)    & 3.106(2)      \\
\ce{C2H5OH}     &	1.466(3)  & 1.479(3)	&	\ce{H2O}	&	4.715(2)    &   4.670(2)	&	\ce{O2+}  	&	1.380(1)	& 1.420(1)	 	\\
\ce{C2H6}       &	1.264(3)  & 1.284(3)	&	\ce{H2O2}	&	3.846(2)    &   3.893(2)	&	\ce{O3}  	&	5.534(2)    & 5.633(2)      \\	
\ce{C3}         &	2.642(3)  & 2.600(3) 	&	\ce{H2S}	&	1.387(3)    &   1.407(3)	&	\ce{OCS}  	&	2.319(3)    & 2.260(3)  	\\
\ce{C3H3}       &	2.762(1)  & 2.765(1)	&	\ce{H3+}	&	2.295($-$1)	&	1.746($-$1) &	\ce{OH}   	&	2.646(2)    & 2.681(2)  	\\
\ce{C3H7OH}     &	2.434(3)  & 2.470(3)	&	\ce{HC3H}	&	3.659(2)    &   3.708(2)	&	\ce{OH+}	&	8.375		& 8.063	    \\
\ce{C4H-}       &	--		  &	--        	&	\ce{HC3N}	&	2.909(3)    &   2.901(3)	&	\ce{P}   	&	--		    & --      		\\
\ce{C6H-}       &	--		  &	--        	&	\ce{HCN}	&	1.052(3)    &   1.069(3)	&	\ce{PH}	    &	3.450(2)	& 3.318(2)	   	\\
\ce{CH}         &	4.598(2)  & 4.468(2)	&	\ce{HCO}	&	2.211(2)    &   2.110(2)	&	\ce{PH+}	&	5.632(1)	& 5.736(1)		\\
\ce{CH+}        &	2.045(2)  &	1.927(2)    &	\ce{HCO+}	&	3.049		&	3.770      	&	\ce{Rb}  	&	--		    & --     		\\
\ce{CH2}        &	1.398(2)  &	1.383(2)   	&	\ce{HCOOH}	&	9.896(2)    &   1.006(3)	&	\ce{S}    	&	--	      	& --     		\\
\ce{CH2+}       &	7.252(1)  &	7.117(1)    &	\ce{HCl}	&	1.090(3)    &   9.841(2)	&	\ce{S2}    	&	3.176(1)    & 3.168(1)  	\\
\ce{CH3}        &	9.999(1)  & 1.058(2) 	&	\ce{HCl+}	&	6.558(1)	&	6.413(1) 	&	\ce{SH}  	&	3.782(2)    & 3.720(2)  	\\
\ce{CH3CHO}     &	8.908(2)  & 8.989(2)	&	\ce{HF}	    &	1.022(2)    &   1.031(2)	&	\ce{SH+}	&	4.309(2)    & 4.220(2)		\\
\ce{CH3CN}      &	1.955(3)  & 2.006(3)	&	\ce{HNC}	&	9.632(2)    &   9.729(2)	&	\ce{SO}   	&	2.791(3)    & 2.827(3)      \\	
\ce{CH3NH2} 	&	1.777(2)  & 1.755(2)	&	\ce{HNCO}	&	1.431(3)    &   1.449(3)	&	\ce{SO2}	&	1.528(3)    & 1.534(3)      \\	
\ce{CH3OCH3}	&	9.651(2)  & 9.637(2)	&	\ce{HO2}	&	8.162(1)    &   8.122(1)	&	\ce{Si}  	&	--		    & --       		\\
\ce{CH3OCHO}	&	1.171(3)  & 1.177(3)	&	\ce{K}	    &	--		    &	--         	&	\ce{SiH}	&	1.852(2)    & 1.867(2)  	\\
\ce{CH3OH}      &	8.862(2)  & 8.892(2)	&	\ce{Li}	    &	--		    &	--         	&	\ce{SiH+}	&	6.804(2)	& 6.981(2)	 	\\
\ce{CH3SH}      &	1.197(3)  & 1.203(3)	&	\ce{LiH}	&	1.522(2)	&	1.518(2)	&	\ce{SiO}	&	5.673(2)	& 5.855(2)	 	\\
\ce{CH4}        &	9.917(2)  & 9.972(2)	&	\ce{Mg}    	&	--		    &	--          &	\ce{Ti} 	&	--		    & --      		\\
\ce{CH4+}       &	1.852(2)  &	1.840(2)    &	\ce{MgH}	&	8.085(1)	&	9.088(1)    &	\ce{Zn} 	&	--		    & --     		\\
\ce{CN}         &	4.287(2)  & 4.350(2)	&	\ce{Mn}	    &	--		    &	--         	&	\ce{c-C3H}	&	1.668(2)    & 1.645(2)      \\	
\ce{CO}         &	1.192(2)  & 6.450(1)	&	\ce{N}	    &	--		    &	--         	&	\ce{c-C3H2}	&	2.523(2)    & 2.554(2)      \\
\ce{CO+}        &	4.960(1)  &	5.239(1)    &	\ce{N2} 	&	3.372(1)    &   4.319(1)	&	\ce{l-C3H}	&	1.326(3)    & 1.281(3)  	\\
\ce{CO2}        &	5.850(2)  & 6.443(2)	&	\ce{N2O}	&	9.807(2)    &   9.976(2)	&	\ce{l-C3H2}	&	1.454(3)	& 1.432(3)	   	\\
\ce{CS}         &	6.416(2)  & 5.137(2)	&	\ce{NH}    	&	1.932(2)    &   1.860(2)	&	\ce{l-C4}	&	4.637(2)    & 4.654(2)	  	\\
\ce{CS2}        &	2.868(3)  & 2.687(3)	&	\ce{NH+}	&	1.678(1)	&	1.647(1)    &	\ce{l-C4H}	&	2.321(3)	& 2.314(3)	    \\
\ce{Ca}         &	--		  &	--        	&	\ce{NH2}	&	3.539(2)    &   3.559(2)	&	\ce{l-C5H}	&	4.787(1)	& 4.790(1)	  	\\
\bottomrule
\end{tabular}
}
\tablefoot{The photodissociation rate is given by $k^{\rm pd}(N)=k_{0,s}^{\rm pd}\zeta_{\ce{H2}}(N)$.
Parameterisations for $\zeta_{\ce{H2}}(N)$ are given in Table~\ref{tab:zetafit}.
Numbers in brackets indicate the power of ten, namely $m(n) = m \times 10^n$.}
\end{center}
\label{tab:pd3.1}
\end{table*}%

\begin{table*}[!h]
\caption{Photoionisation rates normalised to the CR ionisation rate, $k_{0,s}^{\rm pi}$, 
for \ce{H2} o:p=0:1 and \ce{H2} o:p=1:0 and for $R_V=3.1$.}
\begin{center}
\resizebox{\linewidth}{!}{
\begin{tabular}{lcc|lcc|lcc}
\toprule\toprule
\tcbox[sharp corners, colframe=black, colback=white, size=fbox]{$R_V=3.1$}& \multicolumn{2}{c}{$k_{0,s}^{\rm pi}$} && \multicolumn{2}{c}{$k_{0,s}^{\rm pi}$} && \multicolumn{2}{c}{$k_{0,s}^{\rm pi}$}\\
\cmidrule{2-3} \cmidrule{5-6} \cmidrule{8-9}
Species $s$ & o:p=0:1 & o:p=1:0 & Species $s$ & o:p=0:1 & o:p=1:0 & Species $s$ & o:p=0:1 & o:p=1:0\\
\midrule
\ce{Al}         &	1.023(3)  &	9.708(2)	&	\ce{Ca+}	&	1.585		& 1.465	      &	\ce{NH2CHO}	&	3.739(2)    & 3.681(2)      \\
\ce{AlH}        &	6.841(1)  & 6.918(1)	&	\ce{Cl}  	&	5.409(1)	& 5.322(1)	  &	\ce{NH3}	&	1.887(2)    & 1.896(2)  	\\
\ce{C}          &	2.333(2)  &	2.381(2)	&	\ce{Cr} 	&	4.909(2)	& 4.967(2)	  &	\ce{NO}	    &	1.793(2)    & 1.791(2)  	\\
\ce{C2}         &	2.193(2)  & 2.055(2)	&	\ce{Fe}    	&	2.064(2)	& 2.093(2)    &	\ce{NO2}	&	1.032(2)    & 1.009(2)  	\\
\ce{C2H}        &	4.367(2)  & 4.197(2)	&	\ce{H}	    &	4.078		& 3.615 	  &	\ce{Na}    	&	6.693		& 6.704	        \\
\ce{C2H-}       &	6.119(2)  &	6.107(2)	&	\ce{H-}    	&	9.240(2)	& 9.247(2)	  &	\ce{NaCl}	&	--          & --            \\	
\ce{C2H2}       &	3.733(2)  & 3.603(2)	&	\ce{H2}	    &	5.451($-$2)	& 5.717($-$2) &	\ce{NaH}	&	--      	& --     	    \\	
\ce{C2H3}       &	1.335(3)  & 1.349(3)    &	\ce{H2+}	&	--		    & --          &	\ce{Ni}    	&	4.789(1)    & 4.887(1)		\\
\ce{C2H4}       &	2.668(2)  & 2.617(2)	&	\ce{H2CO}	&	2.703(2)	& 2.683(2)    &	\ce{O}   	&	2.668	    & 2.358 	    \\
\ce{C2H5}       &	2.401(2)  & 2.425(2)	&	\ce{H2CS}	&	1.066(3)	& 1.065(3)    &	\ce{O2}  	&	3.338(1)    & 2.656(1)      \\
\ce{C2H5OH}     &	3.620(2)  & 3.484(2)	&	\ce{H2O}	&	2.289(1)	& 2.106(1)    &	\ce{O2+}  	&	--      	& --     	 	\\
\ce{C2H6}       &	1.827(2)  & 1.687(2)	&	\ce{H2O2}	&	1.804(2)	& 1.730(2)    &	\ce{O3}  	&	3.324(1)    & 3.048(1)      \\	
\ce{C3}         &	9.367(1)  & 9.051(1) 	&	\ce{H2S}	&	5.771(2)	& 5.735(2)    &	\ce{OCS}  	&	5.284(2)    & 5.053(2)  	\\
\ce{C3H3}       &	1.022(3)  & 1.035(3)	&	\ce{H3+}	&	--		    & --          &	\ce{OH}   	&	1.662(1)    & 1.569(1)  	\\
\ce{C3H7OH}     &	5.667(2)  & 5.468(2)	&	\ce{HC3H}	&	1.290(3)	& 1.306(3)    &	\ce{OH+}	&	--	    	& --       	    \\
\ce{C4H-}       &	5.022(2)  &	5.013(2)	&	\ce{HC3N}	&	1.696(2)	& 1.575(2)    &	\ce{P}   	&	1.330(3)    & 1.359(3)		\\
\ce{C6H-}       &	3.042(2)  &	3.036(2)	&	\ce{HCN}	&	1.401(1)	& 1.136(1)    &	\ce{PH}	    &	--      	& --    	   	\\
\ce{CH}         &	1.621(2)  & 1.624(2)	&	\ce{HCO}	&	5.843(2)	& 5.904(2)    &	\ce{PH+}	&	--      	& --    		\\
\ce{CH+}        &	--		  &	--          &	\ce{HCO+}	&	--	    	& --          &	\ce{Rb}  	&	1.095(1)    & 1.102(1)		\\
\ce{CH2}        &	--	      &	--        	&	\ce{HCOOH}	&	2.035(2)	& 1.966(2)    &	\ce{S}    	&	7.989(2)  	& 7.278(2)		\\
\ce{CH2+}       &	--		  &	--          &	\ce{HCl}	&	4.771(1)	& 4.243(1)    &	\ce{S2}    	&	1.667(3)    & 1.699(3)  	\\
\ce{CH3}        &	2.433(2)  & 2.458(2) 	&	\ce{HCl+}	&	--	     	& --          &	\ce{SH}  	&	9.110(2)    & 9.244(2)  	\\
\ce{CH3CHO}     &	6.120(2)  & 6.256(2)	&	\ce{HF}	    &	1.026($-$2)	& 9.336($-$3) &	\ce{SH+}	&	--          & --      		\\
\ce{CH3CN}      &	9.682(1)  & 8.834(1)	&	\ce{HNC}	&	1.518(2)	& 1.370(2)    &	\ce{SO}   	&	4.034(2)    & 4.298(2)      \\	
\ce{CH3NH2} 	&	1.157(3)  & 1.160(3)	&	\ce{HNCO}	&	7.642(1)	& 7.172(1)    &	\ce{SO2}	&	1.056(2)    & 9.662(1)      \\	
\ce{CH3OCH3}	&	8.670(2)  & 8.654(2)	&	\ce{HO2}	&	5.597(1)	& 5.442(1)    &	\ce{Si}  	&	2.030(3)    & 2.000(3)		\\
\ce{CH3OCHO}	&	2.178(2)  & 2.114(2)	&	\ce{K}	    &	1.608(1)	& 1.606(1)    &	\ce{SiH}	&	3.141(3)    & 3.146(3)  	\\
\ce{CH3OH}      &	2.314(2)  & 2.256(2)	&	\ce{Li}	    &	1.130(2)	& 1.126(2)    &	\ce{SiH+}	&	--      	& --       	 	\\
\ce{CH3SH}      &	1.354(3)  & 1.379(3)	&	\ce{LiH}	&	--    		& --          &	\ce{SiO}	&	--      	& --      	 	\\
\ce{CH4}        &	2.330(1)  & 2.039(1)	&	\ce{Mg}    	&	4.863(1)    & 4.797(1)    &	\ce{Ti} 	&	1.299(2)    & 1.312(2)		\\
\ce{CH4+}       &	--		  &	--          &	\ce{MgH}	&	--		    & --          &	\ce{Zn} 	&	4.198(2)    & 3.045(2)		\\
\ce{CN}         &	7.371	  & 5.584	    &	\ce{Mn}	    &	1.444(1)	& 1.453(1)    &	\ce{c-C3H}	&	9.757(2)    & 9.857(2)      \\	
\ce{CO}         &	1.438(1)  & 1.029(1)	&	\ce{N}	    &	9.865($-$1)	& 8.447($-$1) &	\ce{c-C3H2}	&	1.227(3)    & 1.243(3)      \\
\ce{CO+}        &	--		  &	--          &	\ce{N2} 	&	1.759($-$1)	& 1.312($-$1) &	\ce{l-C3H}	&	9.998(2)    & 1.012(3)  	\\
\ce{CO2}        &	1.144(1)  & 9.550	    &	\ce{N2O}	&	1.079(2)	& 1.061(2)    &	\ce{l-C3H2}	&	--      	& --       	   	\\
\ce{CS}         &	6.684(1)  & 6.861(1)	&	\ce{NH}    	&	7.819		& 6.735       &	\ce{l-C4}	&	--          & --       	   	\\
\ce{CS2}        &	2.770(2)  & 2.782(2)	&	\ce{NH+}	&	--		    & --          &	\ce{l-C4H}	&	--	    	& --       	   	\\
\ce{Ca}         &	1.116(2)  &	1.110(2)	&	\ce{NH2}	&	6.265(1)	& 6.122(1)    &	\ce{l-C5H}	&	--      	& --       	   	\\
\bottomrule
\end{tabular}
}
\tablefoot{The photoionisation rate is given by $k^{\rm pi}(N)=k_{0,s}^{\rm pi}\zeta_{\ce{H2}}(N)$.
Parameterisations for $\zeta_{\ce{H2}}(N)$ are given in Table~\ref{tab:zetafit}.
Numbers in brackets indicate the power of ten, namely $m(n) = m \times 10^n$.}
\end{center}
\label{tab:pi3.1}
\end{table*}%

\begin{table*}[!h]
\caption{Same as Table~\ref{tab:pd3.1}, but for $R_V=4.0$.}
\begin{center}
\resizebox{\linewidth}{!}{
\begin{tabular}{lcc|lcc|lcc}
\toprule\toprule
\tcbox[sharp corners, colframe=black, colback=white, size=fbox]{$R_V=4.0$} & \multicolumn{2}{c}{$k_{0,s}^{\rm pd}$} & & \multicolumn{2}{c}{$k_{0,s}^{\rm pd}$} && \multicolumn{2}{c}{$k_{0,s}^{\rm pd}$}\\
\cmidrule{2-3} \cmidrule{5-6} \cmidrule{8-9}
Species $s$ & o:p=0:1 & o:p=1:0 & Species $s$ & o:p=0:1 & o:p=1:0 & Species $s$ & o:p=0:1 & o:p=1:0\\
\midrule
\ce{Al}         &	--		  &	--	        &	\ce{Ca+}	&	--		    & --      	&	\ce{NH2CHO}	&	1.371(3)    & 1.374(3)      \\
\ce{AlH}        &	1.359(1)  & 1.361(1)	&	\ce{Cl}  	&	--		    & --      	&	\ce{NH3}	&	6.968(2)    & 6.780(2)  	\\
\ce{C}          &	--	      &	--	        &	\ce{Cr} 	&	--		    & --      	&	\ce{NO}	    &	1.959(2)    & 1.913(2)  	\\
\ce{C2}         &	1.622(2)  & 1.620(2)	&	\ce{Fe}    	&	--		    & --      	&	\ce{NO2}	&	6.576(2)    & 6.598(2)  	\\
\ce{C2H}        &	8.901(2)  & 9.808(2)	&	\ce{H}	    &	--		    & --      	&	\ce{Na}    	&	--	    	& --       	    \\
\ce{C2H-}       &	--		  &	--        	&	\ce{H-}    	&	--	    	& --      	&	\ce{NaCl}	&	4.562(1)    & 4.568(1)	    \\	
\ce{C2H2}       &	1.174(3)  & 9.373(2)	&	\ce{H2}	    &	4.609(2)    & 4.417(2)  &	\ce{NaH}	&	2.189(2)	& 2.237(2)	    \\	
\ce{C2H3}       &	1.353(2)  & 1.352(2)    &	\ce{H2+}	&	3.614(2)	& 3.673(2)	&	\ce{Ni}    	&	--		    & --     		\\
\ce{C2H4}       &	1.581(3)  & 1.581(3)	&	\ce{H2CO}	&	6.869(2)    & 6.664(2)  &	\ce{O}   	&	--		    & --     	    \\
\ce{C2H5}       &	2.202(2)  & 2.185(2)	&	\ce{H2CS}	&	1.063(3)    & 1.063(3)  &	\ce{O2}  	&	3.293(2)    & 3.323(2)      \\
\ce{C2H5OH}     &	1.635(3)  & 1.650(3)	&	\ce{H2O}	&	5.237(2)    & 5.182(2)  &	\ce{O2+}  	&	1.440(1)	& 1.485(1)	 	\\
\ce{C2H6}       &	1.428(3)  & 1.451(3)	&	\ce{H2O2}	&	4.179(2)    & 4.233(2)  &	\ce{O3}  	&	6.035(2)    & 6.144(2)      \\	
\ce{C3}         &	2.764(3)  & 2.721(3) 	&	\ce{H2S}	&	1.528(3)    & 1.551(3)  &	\ce{OCS}  	&	2.487(3)    & 2.424(3)  	\\
\ce{C3H3}       &	2.269(1)  & 2.271(1)	&	\ce{H3+}	&	2.627($-$1)	&1.996($-$1)&	\ce{OH}   	&	2.896(2)    & 2.937(2)  	\\
\ce{C3H7OH}     &	2.692(3)  & 2.733(3)	&	\ce{HC3H}	&	3.557(2)    & 3.605(2)  &	\ce{OH+}	&	1.030(1)	& 9.962  	    \\
\ce{C4H-}       &	--		  &	--	        &	\ce{HC3N}	&	3.197(3)    & 3.190(3)  &	\ce{P}   	&	--		    & --       		\\
\ce{C6H-}       &	--		  &	--	        &	\ce{HCN}	&	1.189(3)    & 1.207(3)  &	\ce{PH}	    &	3.658(2)	& 3.516(2)	   	\\
\ce{CH}         &	4.776(2)  & 4.639(2)	&	\ce{HCO}	&	2.300(2)    & 2.193(2)  &	\ce{PH+}	&	6.336(1)	& 6.455(1)		\\
\ce{CH+}        &	2.497(2)  &	2.361(2)    &	\ce{HCO+}	&	3.612		& 4.467 	&	\ce{Rb}  	&	--		    & --      		\\
\ce{CH2}        &	1.442(2)  &	1.426(2)   	&	\ce{HCOOH}	&	1.086(3)    & 1.106(3)  &	\ce{S}    	&	--	      	& --      		\\
\ce{CH2+}       &	8.697(1)  &	8.563(1)    &	\ce{HCl}	&	1.240(3)    & 1.121(3)  &	\ce{S2}    	&	2.870(1)    & 2.862(1)   	\\
\ce{CH3}        &	1.068(2)  & 1.131(2) 	&	\ce{HCl+}	&	7.726(1)	& 7.569(1)	&	\ce{SH}  	&	3.956(2)    & 3.888(2)   	\\
\ce{CH3CHO}     &	9.463(2)  & 9.556(2)	&	\ce{HF}	    &	1.151(2)    & 1.162(2)  &	\ce{SH+}	&	5.163(2)    & 5.062(2)		\\
\ce{CH3CN}      &	2.230(3)  & 2.288(3)	&	\ce{HNC}	&	1.074(3)    & 1.083(3)  &	\ce{SO}   	&	3.042(3)    & 3.082(3)      \\	
\ce{CH3NH2} 	&	1.847(2)  & 1.824(2)	&	\ce{HNCO}	&	1.583(3)    & 1.604(3)  &	\ce{SO2}	&	1.698(3)    & 1.705(3)      \\	
\ce{CH3OCH3}	&	1.071(3)  & 1.070(3)	&	\ce{HO2}	&	8.531(1)    & 8.467(1)  &	\ce{Si}  	&	--		    & --     		\\
\ce{CH3OCHO}	&	1.261(3)  & 1.269(3)	&	\ce{K}	    &	--		    & --     	&	\ce{SiH}	&	1.732(2)    & 1.748(2)  	\\
\ce{CH3OH}      &	9.874(2)  & 9.915(2)	&	\ce{Li}	    &	--		    & --     	&	\ce{SiH+}	&	7.474(2)	& 7.673(2)	 	\\
\ce{CH3SH}      &	1.283(3)  & 1.289(3)	&	\ce{LiH}	&	1.392(2)	& 1.388(2)	&	\ce{SiO}	&	6.452(2)	& 6.681(2)	 	\\
\ce{CH4}        &	1.135(3)  & 1.141(3)  	&	\ce{Mg}    	&	--		    & --       	&	\ce{Ti} 	&	--		    & --      		\\
\ce{CH4+}       &	2.069(2)  &	2.052(2)    &	\ce{MgH}	&	8.493(1)	& 9.560(1)	&	\ce{Zn} 	&	--		    & --       		\\
\ce{CN}         &	4.992(2)  & 5.079(2)	&	\ce{Mn}	    &	--		    & --	    &	\ce{c-C3H}	&	1.682(2)    & 1.658(2)      \\	
\ce{CO}         &	1.348(2)  & 7.593(1)	&	\ce{N}	    &	--		    & --	    &	\ce{c-C3H2}	&	2.638(2)    & 2.669(2)      \\
\ce{CO+}        &	5.509(1)  &	5.814(1)    &	\ce{N2} 	&	3.921(1)    & 4.679(1)  &	\ce{l-C3H}	&	1.359(3)    & 1.313(3)  	\\
\ce{CO2}        &	6.932(2)  & 7.631(2)	&	\ce{N2O}	&	1.098(3)    & 1.118(3)  &	\ce{l-C3H2}	&	1.551(3)	& 1.528(3)	   	\\
\ce{CS}         &	6.976(2)  & 5.576(2)	&	\ce{NH}    	&	2.187(2)    & 2.105(2)  &	\ce{l-C4}	&	4.224(2)    & 4.239(2)	  	\\
\ce{CS2}        &	3.110(3)  & 2.927(3)	&	\ce{NH+}	&	1.972(1)	& 1.938(1)	&	\ce{l-C4H}	&	2.377(3)	& 2.371(3)	    \\
\ce{Ca}         &	--		  &	--	        &	\ce{NH2}	&	3.732(2)    & 3.758(2)  &	\ce{l-C5H}	&	4.208(1)	& 4.210(1)	  	\\
\bottomrule
\end{tabular}
}
\end{center}
\label{tab:pd4.0}
\end{table*}%

\begin{table*}[!h]
\caption{Same as Table~\ref{tab:pi3.1}, but for $R_V=4.0$.}
\begin{center}
\resizebox{\linewidth}{!}{
\begin{tabular}{lcc|lcc|lcc}
\toprule\toprule
\tcbox[sharp corners, colframe=black, colback=white, size=fbox]{$R_V=4.0$} & \multicolumn{2}{c}{$k_{0,s}^{\rm pi}$} && \multicolumn{2}{c}{$k_{0,s}^{\rm pi}$} && \multicolumn{2}{c}{$k_{0,s}^{\rm pi}$}\\
\cmidrule{2-3} \cmidrule{5-6} \cmidrule{8-9}
Species $s$ & o:p=0:1 & o:p=1:0 & Species $s$ & o:p=0:1 & o:p=1:0 & Species $s$ & o:p=0:1 & o:p=1:0\\
\midrule
\ce{Al}         &	1.053(3)  &	9.961(2)	&	\ce{Ca+}	&	1.948		& 1.810	    &	\ce{NH2CHO}	&	4.418(2)    & 4.357(2)      \\
\ce{AlH}        &	7.480(1)  & 7.565(1)	&	\ce{Cl}  	&	6.726(1)	& 6.737(1)	&	\ce{NH3}	&	2.194(2)    & 2.205(2)   	\\
\ce{C}          &	2.789(2)  &	2.846(2)	&	\ce{Cr} 	&	5.332(2)	& 5.394(2)	&	\ce{NO}	    &	2.098(2)    & 2.099(2)  	\\
\ce{C2}         &	2.720(2)  & 2.565(2)	&	\ce{Fe}    	&	2.268(2)	& 2.299(2)	&	\ce{NO2}	&	1.238(2)    & 1.212(2)  	\\
\ce{C2H}        &	5.336(2)  & 5.148(2)	&	\ce{H}	    &	4.804		& 4.292    	&	\ce{Na}    	&	7.395		& 7.411 	    \\
\ce{C2H-}       &	6.534(2)  &	6.524(2)	&	\ce{H-}    	&	9.110(2)	& 9.122(2)	&	\ce{NaCl}	&	--          & --       	    \\	
\ce{C2H2}       &	4.535(2)  & 4.392(2)	&	\ce{H2}	    &	5.939($-$2)	& 6.239($-$2)&	\ce{NaH}	&	--      	& --      	    \\	
\ce{C2H3}       &	1.529(3)  & 1.545(3)    &	\ce{H2+}	&	--		    & --     	&	\ce{Ni}    	&	5.050(1)    & 5.152(1)		\\
\ce{C2H4}       &	3.175(2)  & 3.117(2)	&	\ce{H2CO}	&	3.214(2)	& 3.195(2)  &	\ce{O}   	&	3.139	    & 2.786   	    \\
\ce{C2H5}       &	2.706(2)  & 2.734(2)	&	\ce{H2CS}	&	1.220(3)	& 1.218(3)  &	\ce{O2}  	&	4.130(1)    & 3.332(1)      \\
\ce{C2H5OH}     &	4.336(2)  & 4.184(2)	&	\ce{H2O}	&	2.840(1)	& 2.637(1)  &	\ce{O2+}  	&	--      	& --       	 	\\
\ce{C2H6}       &	2.249(2)  & 2.094(2)	&	\ce{H2O2}	&	2.189(2)	& 2.107(2)  &	\ce{O3}  	&	4.109(1)    & 3.807(1)      \\	
\ce{C3}         &	1.154(2)  & 1.120(2) 	&	\ce{H2S}	&	6.771(2)	& 6.738(2)  &	\ce{OCS}  	&	6.375(2)    & 6.117(2)  	\\
\ce{C3H3}       &	1.150(3)  & 1.166(3)	&	\ce{H3+}	&	--		    & --    	&	\ce{OH}   	&	2.053(1)    & 1.967(1)  	\\
\ce{C3H7OH}     &	6.789(2)  & 6.568(2)	&	\ce{HC3H}	&	1.453(3)	& 1.471(3)  &	\ce{OH+}	&	--	    	& --     	    \\
\ce{C4H-}       &	5.382(2)  &	5.375(2)	&	\ce{HC3N}	&	2.071(2)	& 1.934(2)  &	\ce{P}   	&	1.538(3)    & 1.570(3)		\\
\ce{C6H-}       &	3.265(2)  &	3.261(2)	&	\ce{HCN}	&	1.641(1)	& 1.337(1)  &	\ce{PH}	    &	--      	& --      	   	\\
\ce{CH}         &	1.929(2)  & 1.934(2)	&	\ce{HCO}	&	6.736(2)	& 6.807(2)  &	\ce{PH+}	&	--      	& --      		\\
\ce{CH+}        &	--		  &	--          &	\ce{HCO+}	&	--	    	& --    	&	\ce{Rb}  	&	1.186(1)    & 1.193(1)		\\
\ce{CH2}        &	--	      &	--         	&	\ce{HCOOH}	&	2.469(2)    & 2.394(2)  &	\ce{S}    	&	9.387(2)  	& 8.552(2)		\\
\ce{CH2+}       &	--		  &	--          &	\ce{HCl}	&	5.844(1)	& 5.239(1)  &	\ce{S2}    	&	1.883(3)    & 1.919(3)  	\\
\ce{CH3}        &	2.770(2)  & 2.799(2) 	&	\ce{HCl+}	&	--	     	& --     	&	\ce{SH}  	&	1.060(3)    & 1.075(3)  	\\
\ce{CH3CHO}     &	6.923(2)  & 7.081(2)	&	\ce{HF}	    &	1.108($-$2)	& 1.007($-$2)&	\ce{SH+}	&	--          & --       		\\
\ce{CH3CN}      &	1.194(2)  & 1.099(2)	&	\ce{HNC}	&	1.856(2)	& 1.686(2)  &	\ce{SO}   	&	4.658(2)    & 4.975(2)      \\	
\ce{CH3NH2} 	&	1.322(3)  & 1.326(3)	&	\ce{HNCO}	&	9.352(1)	& 8.823(1)  &	\ce{SO2}	&	1.306(2)    & 1.205(2)      \\	
\ce{CH3OCH3}	&	9.970(2)  & 9.954(2)	&	\ce{HO2}	&	6.813(1)	& 6.643(1)  &	\ce{Si}  	&	2.231(3)    & 2.200(3)		\\
\ce{CH3OCHO}	&	2.619(2)  & 2.550(2)	&	\ce{K}	    &	1.741(1)	& 1.740(1)	&	\ce{SiH}	&	3.564(3)    & 3.572(3)  	\\
\ce{CH3OH}      &	2.779(2)  & 2.717(2)	&	\ce{Li}	    &	1.202(2)    & 1.199(2)	&	\ce{SiH+}	&	--      	& --       	 	\\
\ce{CH3SH}      &	1.534(3)  & 1.563(3)	&	\ce{LiH}	&	--    		& --      	&	\ce{SiO}	&	--      	& --       	 	\\
\ce{CH4}        &	2.804(1)  & 2.484(1)	&	\ce{Mg}    	&	5.105(1)    & 5.035(1)	&	\ce{Ti} 	&	1.461(2)    & 1.476(2)		\\
\ce{CH4+}       &	--		  &	--          &	\ce{MgH}	&	--		    & --     	&	\ce{Zn} 	&	5.035(2)    & 3.639(2)		\\
\ce{CN}         &	8.578	  & 6.512   	&	\ce{Mn}	    &	1.499(1)    & 1.509(1)	&	\ce{c-C3H}	&	1.109(3)    & 1.120(3)      \\	
\ce{CO}         &	1.675(1)  & 1.199(1)	&	\ce{N}	    &	1.113    	& 9.547($-$1)&	\ce{c-C3H2}	&	1.384(3)    & 1.403(3)      \\
\ce{CO+}        &	--		  &	--     	    &	\ce{N2} 	&	1.899($-$1)	& 1.426($-$1)&	\ce{l-C3H}	&	1.136(3)    & 1.149(3)   	\\
\ce{CO2}        &	1.341(1)  & 1.124(1)	&	\ce{N2O}	&	1.385(2)	& 1.373(2)  &	\ce{l-C3H2}	&	--      	& --         	\\
\ce{CS}         &	8.054(1)  & 8.263(1)	&	\ce{NH}    	&	9.390		& 8.174     &	\ce{l-C4}	&	--          & --          	\\
\ce{CS2}        &	3.258(2)  & 3.273(2)	&	\ce{NH+}	&	--		    & --    	&	\ce{l-C4H}	&	--	    	& --            \\
\ce{Ca}         &	1.169(2)  &	1.163(2)	&	\ce{NH2}	&	7.638(1)	& 7.490(1)  &	\ce{l-C5H}	&	--      	& --       	  	\\
\bottomrule
\end{tabular}
}
\end{center}
\label{tab:pi4.0}
\end{table*}%

\begin{table*}[!h]
\caption{Same as Table~\ref{tab:pd3.1}, but for $R_V=5.5$.}
\begin{center}
\resizebox{\linewidth}{!}{
\begin{tabular}{lcc|lcc|lcc}
\toprule\toprule
\tcbox[sharp corners, colframe=black, colback=white, size=fbox]{$R_V=5.5$} & \multicolumn{2}{c}{$k_{0,s}^{\rm pd}$} & & \multicolumn{2}{c}{$k_{0,s}^{\rm pd}$} && \multicolumn{2}{c}{$k_{0,s}^{\rm pd}$}\\
\cmidrule{2-3} \cmidrule{5-6} \cmidrule{8-9}
Species $s$ & o:p=0:1 & o:p=1:0 & Species $s$ & o:p=0:1 & o:p=1:0 & Species $s$ & o:p=0:1 & o:p=1:0\\
\midrule
\ce{Al}         &	--		  &	--	        &	\ce{Ca+}	&	--		    & --      	&	\ce{NH2CHO}	&	1.628(3)    & 1.633(3)      \\
\ce{AlH}        &	1.495(1)  & 1.497(1)	&	\ce{Cl}  	&	--		    & --       	&	\ce{NH3}	&	8.540(2)    & 8.318(2)  	\\
\ce{C}          &	--	      &	--         	&	\ce{Cr} 	&	--		    & --      	&	\ce{NO}	    &	2.363(2)    & 2.314(2)      \\
\ce{C2}         &	2.011(2)  & 2.010(2)	&	\ce{Fe}    	&	--		    & --     	&	\ce{NO2}	&	7.991(2)    & 8.017(2)  	\\
\ce{C2H}        &	1.088(3)  & 1.202(3)	&	\ce{H}	    &	--		    & --      	&	\ce{Na}    	&	--	    	& --       	    \\
\ce{C2H-}       &	--		  &	--        	&	\ce{H-}    	&	--	    	& --       	&	\ce{NaCl}	&	5.056(1)    & 5.063(1)	    \\	
\ce{C2H2}       &	1.410(3)  & 1.121(3)	&	\ce{H2}	    &	5.175(2)    & 4.953(2)  &	\ce{NaH}	&	2.411(2)	& 2.469(2)	    \\	
\ce{C2H3}       &	1.498(2)  & 1.497(2)    &	\ce{H2+}	&	4.435(2)	& 4.512(2)	&	\ce{Ni}    	&	--		    & --     		\\
\ce{C2H4}       &	1.901(3)  & 1.903(3)	&	\ce{H2CO}	&	8.342(2)    & 8.101(2)  &	\ce{O}   	&	--		    & --     	    \\
\ce{C2H5}       &	2.495(2)  & 2.476(2)	&	\ce{H2CS}	&	1.287(3)    & 1.288(3)  &	\ce{O2}  	&	3.893(2)    & 3.939(2)      \\
\ce{C2H5OH}     &	1.999(3)  & 2.020(3)	&	\ce{H2O}	&	6.336(2)    & 6.281(2)  &	\ce{O2+}  	&	1.660(1)	& 1.717(1)	 	\\
\ce{C2H6}       &	1.762(3)  & 1.792(3)	&	\ce{H2O2}	&	5.089(2)    & 5.156(2)  &	\ce{O3}  	&	7.306(2)    & 7.446(2)      \\	
\ce{C3}         &	3.233(3)  & 3.184(3) 	&	\ce{H2S}	&	1.858(3)    & 1.888(3)  &	\ce{OCS}  	&	2.958(3)    & 2.884(3)  	\\
\ce{C3H3}       &	2.505(1)  & 2.508(1)	&	\ce{H3+}	&	3.040($-$1)	& 2.314($-$1)&	\ce{OH}   	&	3.507(2)    & 3.561(2)      \\
\ce{C3H7OH}     &	3.290(3)  & 3.344(3)	&	\ce{HC3H}	&	4.010(2)    & 4.065(2)  &	\ce{OH+}	&	1.269(1)	& 1.235(1)	    \\
\ce{C4H-}       &	--		  &	--        	&	\ce{HC3N}	&	3.892(3)    & 3.885(3)  &	\ce{P}   	&	--		    & --       		\\
\ce{C6H-}       &	--		  &	--        	&	\ce{HCN}	&	1.462(3)    & 1.487(3)  &	\ce{PH}	    &	4.361(2)	& 4.188(2)	   	\\
\ce{CH}         &	5.571(2)  & 5.409(2)	&	\ce{HCO}	&	2.717(2)    & 2.590(2)  &	\ce{PH+}	&	7.810(1)	& 7.966(1)		\\
\ce{CH+}        &	3.077(2)  &	2.924(2)    &	\ce{HCO+}	&	4.526		& 5.602  	&	\ce{Rb}  	&	--		    & --     		\\
\ce{CH2}        &	1.704(2)  &	1.686(2)   	&	\ce{HCOOH}	&	1.316(3)    & 1.341(3)  &	\ce{S}    	&	--	      	& --      		\\
\ce{CH2+}       &	1.063(2)  &	1.053(2)    &	\ce{HCl}	&	1.531(3)    & 1.387(3)  &	\ce{S2}    	&	3.120(1)    & 3.112(1)   	\\
\ce{CH3}        &	1.279(2)  & 1.353(2) 	&	\ce{HCl+}	&	9.453(1)	& 9.294(1)	&	\ce{SH}  	&	4.668(2)    & 4.588(2)  	\\
\ce{CH3CHO}     &	1.127(3)  & 1.139(3)	&	\ce{HF}	    &	1.411(2)    & 1.426(2)  &	\ce{SH+}	&	6.381(2)    & 6.279(2)		\\
\ce{CH3CN}      &	2.759(3)  & 2.835(3)	&	\ce{HNC}	&	1.313(3)    & 1.326(3)  &	\ce{SO}   	&	3.734(3)    & 3.783(3)      \\	
\ce{CH3NH2} 	&	2.174(2)  & 2.147(2)	&	\ce{HNCO}	&	1.940(3)    & 1.967(3)  &	\ce{SO2}	&	2.084(3)    & 2.094(3)      \\	
\ce{CH3OCH3}	&	1.301(3)  & 1.301(3)	&	\ce{HO2}	&	1.018(2)    & 1.011(2)  &	\ce{Si}  	&	--		    & --     		\\
\ce{CH3OCHO}	&	1.522(3)  & 1.532(3)	&	\ce{K}	    &	--		    & --     	&	\ce{SiH}	&	1.946(2)    & 1.964(2)   	\\
\ce{CH3OH}      &	1.202(3)  & 1.209(3)	&	\ce{Li}	    &	--		    & --     	&	\ce{SiH+}	&	9.179(2)	& 9.427(2)	 	\\
\ce{CH3SH}      &	1.541(3)  & 1.550(3)	&	\ce{LiH}	&	1.515(2)	& 1.511(2)	&	\ce{SiO}	&	7.939(2)	& 8.243(2)	 	\\
\ce{CH4}        &	1.398(3)  & 1.409(3)	&	\ce{Mg}    	&	--		    & --     	&	\ce{Ti} 	&	--		    & --      		\\
\ce{CH4+}       &	2.535(2)  &	2.518(2)    &	\ce{MgH}	&	9.977(1)	& 1.124(2)	&	\ce{Zn} 	&	--		    & --       		\\
\ce{CN}         &	6.169(2)  & 6.296(2)	&	\ce{Mn}	    &	--		    & --     	&	\ce{c-C3H}	&	1.931(2)    & 1.903(2)      \\	
\ce{CO}         &	1.491(2)  & 8.739(1)	&	\ce{N}	    &	--		    & --      	&	\ce{c-C3H2}	&	3.100(2)    & 3.132(2)      \\
\ce{CO+}        &	6.850(1)  &	7.230(1)    &	\ce{N2} 	&	4.443(1)    & 5.001(1)  &	\ce{l-C3H}	&	1.570(3)    & 1.516(3)  	\\
\ce{CO2}        &	8.542(2)  & 9.430(2)	&	\ce{N2O}	&	1.358(3)    & 1.383(3)  &	\ce{l-C3H2}	&	1.856(3)	& 1.830(3)   	\\
\ce{CS}         &	8.464(2)  & 6.768(2)	&	\ce{NH}    	&	2.669(2)    & 2.573(2)  &	\ce{l-C4}	&	4.642(2)    & 4.658(2)	  	\\
\ce{CS2}        &	3.744(3)  & 3.539(3)	&	\ce{NH+}	&	2.421(1)	& 2.389(1)	&	\ce{l-C4H}	&	2.737(3)	& 2.730(3)	    \\
\ce{Ca}         &	--		  &	--      	&	\ce{NH2}	&	4.432(2)    & 4.466(2)  &	\ce{l-C5H}	&	4.635(1)	& 4.638(1)	  	\\
\bottomrule
\end{tabular}
}
\end{center}
\label{tab:pd5.5}
\end{table*}%

\begin{table*}[!h]
\caption{Same as Table~\ref{tab:pi3.1}, but for $R_V=5.5$.}
\begin{center}
\resizebox{\linewidth}{!}{
\begin{tabular}{lcc|lcc|lcc}
\toprule\toprule
\tcbox[sharp corners, colframe=black, colback=white, size=fbox]{$R_V=5.5$} & \multicolumn{2}{c}{$k_{0,s}^{\rm pi}$} && \multicolumn{2}{c}{$k_{0,s}^{\rm pi}$} && \multicolumn{2}{c}{$k_{0,s}^{\rm pi}$}\\
\cmidrule{2-3} \cmidrule{5-6} \cmidrule{8-9}
Species $s$ & o:p=0:1 & o:p=1:0 & Species $s$ & o:p=0:1 & o:p=1:0 & Species $s$ & o:p=0:1 & o:p=1:0\\
\midrule
\ce{Al}         &	1.238(3)  &	1.168(3)	&	\ce{Ca+}	&	2.403		& 2.245	    &	\ce{NH2CHO}	&	5.442(2)    & 5.388(2)      \\
\ce{AlH}        &	9.150(1)  & 9.256(1)	&	\ce{Cl}  	&	8.103(1)	& 8.225(1)	&	\ce{NH3}	&	2.712(2)    & 2.734(2)  	\\
\ce{C}          &	3.437(2)  &	3.521(2)	&	\ce{Cr} 	&	6.429(2)	& 6.509(2)	&	\ce{NO}	    &	2.585(2)    & 2.594(2)      \\
\ce{C2}         &	3.348(2)  & 3.176(2)	&	\ce{Fe}    	&	2.751(2)	& 2.789(2)	&	\ce{NO2}	&	1.532(2)    & 1.506(2)  	\\
\ce{C2H}        &	6.593(2)  & 6.392(2)	&	\ce{H}	    &	5.538		& 4.966    	&	\ce{Na}    	&	8.980		& 9.013         \\
\ce{C2H-}       &	7.804(2)  &	7.802(2)	&	\ce{H-}    	&	1.014(3)	& 1.016(3)	&	\ce{NaCl}	&	--          & --       	    \\	
\ce{C2H2}       &	5.576(2)  & 5.431(2)	&	\ce{H2}	    &	6.871($-$2)	& 7.230($-$2)&	\ce{NaH}	&	--      	& --      	    \\	
\ce{C2H3}       &	1.885(3)  & 1.909(3)    &	\ce{H2+}	&	--		    & --     	&	\ce{Ni}    	&	5.949(1)    & 6.067(1)		\\
\ce{C2H4}       &	3.928(2)  & 3.870(2)	&	\ce{H2CO}	&	3.966(2)	& 3.958(2)  &	\ce{O}   	&	3.623	    & 3.223   	    \\
\ce{C2H5}       &	3.333(2)  & 3.371(2)	&	\ce{H2CS}	&	1.501(3)	& 1.503(3)  &	\ce{O2}  	&	5.020(1)    & 4.099(1)      \\
\ce{C2H5OH}     &	5.320(2)  & 5.160(2)	&	\ce{H2O}	&	3.425(1)	& 3.204(1)  &	\ce{O2+}  	&	--      	& --       	 	\\
\ce{C2H6}       &	2.731(2)  & 2.560(2)	&	\ce{H2O2}	&	2.684(2)	& 2.597(2)  &	\ce{O3}  	&	4.937(1)    & 4.610(1)      \\	
\ce{C3}         &	1.420(2)  & 1.386(2) 	&	\ce{H2S}	&	8.363(2)	& 8.350(2)  &	\ce{OCS}  	&	7.827(2)    & 7.553(2)   	\\
\ce{C3H3}       &	1.417(3)  & 1.438(3)	&	\ce{H3+}	&	--		    & --       	&	\ce{OH}   	&	2.454(1)    & 2.379(1)      \\
\ce{C3H7OH}     &	8.336(2)  & 8.105(2)	&	\ce{HC3H}	&	1.791(3)	& 1.816(3)  &	\ce{OH+}	&	--	    	& --     	    \\
\ce{C4H-}       &	6.441(2)  &	6.440(2)	&	\ce{HC3N}	&	2.536(2)    & 2.382(2)  &	\ce{P}   	&	1.907(3)    & 1.951(3)		\\
\ce{C6H-}       &	3.910(2)  &	3.910(2)	&	\ce{HCN}	&	1.899(1)	& 1.552(1)  &	\ce{PH}	    &	--      	& --      	   	\\
\ce{CH}         &	2.384(2)  & 2.399(2)	&	\ce{HCO}	&	8.330(2)	& 8.435(2)  &	\ce{PH+}	&	--      	& --     		\\
\ce{CH+}        &	--		  &	--          &	\ce{HCO+}	&	--	    	& --    	&	\ce{Rb}  	&	1.424(1)    & 1.435(1)		\\
\ce{CH2}        &	--	      &	--        	&	\ce{HCOOH}	&	3.033(2)    & 2.957(2)  &	\ce{S}    	&	1.162(3)  	& 1.061(3)		\\
\ce{CH2+}       &	--		  &	--          &	\ce{HCl}	&	6.990(1)	& 6.304(1)  &	\ce{S2}    	&	2.317(3)    & 2.365(3)  	\\
\ce{CH3}        &	3.415(2)  & 3.456(2) 	&	\ce{HCl+}	&	--	     	& --     	&	\ce{SH}  	&	1.311(3)    & 1.333(3)  	\\
\ce{CH3CHO}     &	8.588(2)  & 8.796(2)	&	\ce{HF}	    &	1.287($-$2)	& 1.167($-$2)&	\ce{SH+}	&	--          & --       		\\
\ce{CH3CN}      &	1.441(2)  & 1.336(2)	&	\ce{HNC}	&	2.260(2)	& 2.068(2)  &	\ce{SO}   	&	5.796(2)    & 6.201(2)      \\	
\ce{CH3NH2} 	&	1.628(3)  & 1.636(3)	&	\ce{HNCO}	&	1.146(2)	& 1.087(2)  &	\ce{SO2}	&	1.576(2)    & 1.465(2)      \\	
\ce{CH3OCH3}	&	1.226(3)  & 1.227(3)	&	\ce{HO2}	&	8.429(1)	& 8.260(1)  &	\ce{Si}  	&	2.703(3)    & 2.670(3)		\\
\ce{CH3OCHO}	&	3.216(2)  & 3.148(2)	&	\ce{K}	    &	2.093(1)    & 2.094(1)	&	\ce{SiH}	&	4.380(3)    & 4.397(3)  	\\
\ce{CH3OH}      &	3.414(2)  & 3.355(2)	&	\ce{Li}	    &	1.433(2)    & 1.431(2)	&	\ce{SiH+}	&	--      	& --       	 	\\
\ce{CH3SH}      &	1.895(3)  & 1.934(3)	&	\ce{LiH}	&	--    		& --    	&	\ce{SiO}	&	--      	& --     	 	\\
\ce{CH4}        &	3.304(1)  & 2.954(1)	&	\ce{Mg}    	&	5.994(1)    & 5.913(1)	&	\ce{Ti} 	&	1.798(2)    & 1.818(2)		\\
\ce{CH4+}       &	--		  &	--          &	\ce{MgH}	&	--		    & --     	&	\ce{Zn} 	&	6.334(2)    & 4.563(2)		\\
\ce{CN}         &	9.887	  & 7.517   	&	\ce{Mn}	    &	1.745(1)    & 1.756(1)	&	\ce{c-C3H}	&	1.367(3)    & 1.384(3)      \\	
\ce{CO}         &	1.934(1)  & 1.386(1)	&	\ce{N}	    &	1.293    	& 1.112	    &	\ce{c-C3H2}	&	1.707(3)    & 1.732(3)      \\
\ce{CO+}        &	--		  &	--          &	\ce{N2} 	&	2.169($-$1)	& 1.648($-$1)&	\ce{l-C3H}	&	1.401(3)    & 1.420(3)  	\\
\ce{CO2}        &	1.545(1)  & 1.298(1) 	&	\ce{N2O}	&	1.702(2)	& 1.701(2)  &	\ce{l-C3H2}	&	--      	& --     	   	\\
\ce{CS}         &	9.894(1)  & 1.019(2)	&	\ce{NH}    	&	1.099(1)	& 9.642     &	\ce{l-C4}	&	--          & --      	  	\\
\ce{CS2}        &	4.028(2)  & 4.060(2)	&	\ce{NH+}	&	--		    & --       	&	\ce{l-C4H}	&	--	    	& --     	    \\
\ce{Ca}         &	1.377(2)  &	1.372(2)	&	\ce{NH2}	&	9.378(1)	& 9.252(1)  &	\ce{l-C5H}	&	--      	& --       	  	\\
\bottomrule
\end{tabular}
}
\end{center}
\label{tab:pi5.5}
\end{table*}%

\end{appendix}

\end{document}